\documentclass[prd,twocolumn,amsmath,amssymb,superscriptaddress,floatfix,notitlepage,nofootinbib]{revtex4-1}
\usepackage{hyperref}
\usepackage{psfrag} 

\usepackage{natbib}
\usepackage{graphicx}
\usepackage{txfonts}

\usepackage{longtable}

\usepackage{relsize}
\usepackage{pbox}
\usepackage[usenames,dvipsnames]{color}
\usepackage{bm}

\usepackage{soul}
\soulregister\cite7
\soulregister\ref7
\soulregister\figref7
\soulregister\tabref7
\soulregister\cite7
\soulregister\citep7
\soulregister\citet7

\usepackage{maobsabbrevs}
\usepackage{todonotes}
\usepackage{xspace}

 \newcommand{\bref}[1]{(\ref{#1})}

 \newcommand{\eqsref}[1]{Eqs.\,(\ref{#1})}

 \newcommand{\tto}{\text{--}}

 \newcommand{\der}{\mathrm{d}} 

\newcommand{\pasa}{PASA}

 \newcommand{\jcop}{J. Comput. Phys.}
 \newcommand{\mnras}{MNRAS}

\newcommand{\aap}{A{\&}A}
\renewcommand{\prd}{Phys. Rev. D}
\newcommand{\apjl}{ApJL}

 \newcommand{\apjs}{ApJS}

\newcommand{\tento}[2]{{#1} \cdot 10^{#2}}
\newcommand{\etoten}{ \cdot 10^ }

\newcommand{\kb}{ k_{\rm B}  }
\newcommand{\beq}{\begin{equation}}
\newcommand{\eeq}{\end{equation}}
\newcommand{\beqa}{\begin{eqnarray}}
\newcommand{\eeqa}{\end{eqnarray}}
\newcommand{\beqar}{\begin{eqnarray*}}
\newcommand{\eeqar}{\end{eqnarray*}}

\newcommand{\aenus}{\textsc{Aenus}}

\newcommand{\Qq}{{\cal Q}}

\newcommand{\mev}{\,\rm{MeV}}

 \newcommand{\maa}[1]{\textcolor{Brown}{#1}}

\begin{document}

 \title {Heavy sterile neutrinos in stellar core-collapse}

\author{T.~Rembiasz}
\author{M.~Obergaulinger}
\affiliation{Departamento de Astronom\'{\i}a y Astrof\'{\i}sica,  Universidad de Valencia,  C/ Dr.~Moliner 50, E-46100 Burjassot, Spain}

\author{M.~Masip}
\affiliation{CAFPE and Departamento de F{\'\i}sica Te\'orica y del Cosmos, Universidad de Granada, E-18071 Spain}

\author{M. A.~P\'erez-Garc\'ia}
\affiliation{Department of Fundamental Physics, University of Salamanca, Plaza de la Merced s/n E-37008 Spain}

\author{M.A.~Aloy}
\affiliation{Departamento de Astronom\'{\i}a y Astrof\'{\i}sica,  Universidad de Valencia,  C/ Dr.~Moliner 50, E-46100 Burjassot, Spain}

\author{C.~Albertus}
\affiliation{Department of Fundamental Physics, University of  Salamanca, Plaza de la Merced s/n E-37008 Spain}

\date{Received 31 May 2018; accepted 23 October 2018}


\label{firstpage}

\begin{abstract}
  We perform spherically symmetric simulations of the core collapse of
a single progenitor star of zero age main sequence
mass $M_{\rm ZAMS } = 15 \, M_{\odot}$ with two models of heavy sterile
neutrinos in the mass range of hundred MeV$/c^2$.  According to both
models, these hypothetical particles are copiously produced in the
center, stream outwards a subsequently decay releasing energy into
final states (including neutrinos) of the Standard Model.  We find
that they can lead to a successful explosion in otherwise
non-exploding progenitors.  Depending on their unknown parameters
(e.g., mass and coupling constants with matter), we obtain either no
explosion or an explosion of one of two types, i.e., through heating
of gas downstream of the stalled shock wave, similarly to the standard
scenario for supernova explosions or through heating of gas at higher
radii that ejects matter from the outer core or the envelope while the
center continues to accrete matter.  In both cases, the explosion
energies can be very high.  We presume that this new type of explosion
would produce an electromagnetic signal that significantly differs
from common events because of the relative absence of heavy elements
in the ejecta.  The combination of core-collapse simulations and
astrophysical observations may further constrain the parameters of the
sterile neutrinos.  
\end{abstract}

\maketitle


\section{Introduction} 
\label{sec:intro}

Although neutrinos are a fundamental ingredient of the Standard Model
{(SM)}, it has only been during the last decades that detectors have
reached the sensitivity and statistics required to study their
properties. Today, we have a basic scheme of mass differences and
mixings among the three flavors that provides a good fit to the data,
but some fundamental questions like their absolute mass and hierarchy,
their Dirac or Majorana nature or even the existence of additional
sterile modes remain unanswered. In particular, the presence of some
persistent anomalies in reactor \cite{Mention:2011rk},
Gallium\cite{Giunti:2010zu} and baseline
\cite{Athanassopoulos:1997pv,AguilarArevalo:2008rc,MiniBooNE_2018} experiments
underlines the possibility of a non-minimal neutrino sector
\cite{2018arXiv180303262M}.

Despite their weak couplings, cosmology and astrophysics probe the
properties of neutrinos in a variety of energy ranges. In the early
universe, neutrinos were in thermal equilibrium with matter at the
temperature $T> 1$ MeV/$\kb \approx 10^{10}$ K,%
 \footnote{We use
Heaviside–-Lorentz units with explicitly written $c$, $\hbar,$ and
$\kb$ (the speed of light, the reduced Planck constant, and the
Boltzmann constant, respectively).} 
 whereas in stars like the Sun,
they are copiously produced through nuclear reactions.  Proto-neutron
stars (PNSs) formed during the core-collapse previous to a supernova
explosion are another testing ground for the physics in this sector.
During a $10$--$20$ s period they reach large densities and
temperatures exceeding $\sim$30 MeV/$\kb$, and their long-term
evolution is also sensitive to the presence of any long-lived exotic
particles with sub-GeV$/c^2$ mass
\citep{Raffelt_Zhou_2011,Perez-Garcia_Silk_2015,Cemeno_et_al_2017,
Abazajian+_2001}.  In general, to avoid experimental bounds, the
coupling of these hypothetical particles to matter must be very
weak. This implies that, if produced in a PNS, they tend to escape
faster than standard neutrinos, adding a source of energy loss that
shortens the cooling time and the duration of the neutrino signal from
a supernova explosion.  That argument has been used to set stringent
limits on models with axions, sterile neutrinos or Kaluza-Klein
excitations of the graviton \cite{Hannestad_Raffelt_2001}.

On the other hand, current simulations of supernova explosions seem to
face a generic difficulty. Once the core exhausts all the nuclear fuel
and collapses, most simulations predict the appearance of a stalled
shock-front at a few hundred km from the core. A successful supernova
explosion requires then that a significant fraction of the energy in
the PNS be transferred to the gas behind the shock-front.  The current
standard supernova model is based on the fact that (active) neutrinos
streaming out of the PNS deposit energy in the semitransparent
post-shock layer.  Simulations show that neutrino heating alone does
not suffice to revive the stalled shock in most stars.  Several of
these cases produce an explosion if the efficiency of heating is
enhanced by non-spherical flows.  Currently, open problems include the
conditions for triggering an explosion by this mechanism, the range of
explosion energies and ejecta masses that can be achieved, and whether
or not previous successes from axisymmetric modelling can be
reproduced by the recently started, computationally much more
demanding three-dimensional (3D) simulations (for a review, see e.g.,
\cite{Janka_ARNP_2016,Janka_2017}).  Other mechanisms such as
magnetorotational explosions have been considered
\citep{Mosta_et_al__2014__apjl__MagnetorotationalCore-collapseSupernovaeinThreeDimensions,Obergaulinger_Aloy__2017__mnras__Protomagnetarandblackholeformationinhigh-massstars}.
However, in contrast to neutrino heating, which prevails in any
post-collapse core, they rely on conditions (i.e.~rapid rotation or
strong magnetic fields) that only are present in a small class of
progenitors.

  A parallel line of research considers the potential of variations of
the input from nuclear and particle physics to resolve these open
questions.  Some of these variations involve effects that are
confirmed by particle physics, but whose uncertainties or numerical
complexities so far prevented their implementation in supernova
simulations, like the production of muons \citep{Bollig_2017} or
neutrino flavor oscillations (for a review, see \cite{Mirizzi_2016}).
Others explore more speculative modifications to the standard
microphysics such as the phase transition to quark matter at high
densities, which is capable of producing explosions even in spherical
symmetry \citep{Fischer_2011}.

  In this context, sterile neutrinos \cite{Pontecorvo_1967} are
another interesting possibility. These neutrinos are SM singlets with
neither gauge nor Yukawa interactions with standard quarks and
leptons. However, as we detail in the next section, they may couple to
matter through mixing with the active neutrinos or through one-loop
diagrams involving charged particles in the TeV$/c^2$ mass range.
Sterile neutrinos in the keV$/c^2$ mass range could be copiously
produced in the core. For $m_{\rm{s}}\sim$ 100 keV$/c^2$ (where ``s''
stands for \emph{sterile} neutrino), the vacuum mixing angle of
sterile neutrinos is stringently constrained $\sin^2(2\theta)\lesssim
10^{-9}$ in order to avoid excessive energy loss
\cite{Abazajian+_2001} or to generate supernova asymmetries resulting
in large pulsar kicks \cite{Kusenko_Serge_1997} (on alternative
explanations for neutron star kicks, see, e.g. \cite{Annop_2013}).  In
the range 10 MeV$/c^2 \lesssim m_{\rm{h}} \lesssim m_{K}$ (where ``h''
stands for \emph{heavy} sterile neutrino, and $m_{K}$ is the Kaon
mass), the constraints predominantly come from reactor anomalies and
decays of pions and kaons.  Furthermore, sterile neutrinos have been
considered as a warm dark matter candidate
\cite{Olive_Turner_1982,Dodelson_Widrow_1994,Malaney_et_al_1995,Colombi_et_al_1996,Abazajian+_2001},
or as the origin of the $3.5\,$keV line observed in X-ray telescopes
\cite{Abazajian_2014,Boyarsky_et_al_2014,Bulbul:2014sua}.  Here, we
investigate the possible effects on the dynamics of supernova
explosions and of the remnant PNSs of two such heavy sterile models:
FKP of Fuller et al.~\cite{Fuller+_2009} and AMP of Albertus et
al.~\cite{Albertus+_2015}.  The models include a sterile neutrino
$\nu_{\rm{h}}$ that has a relatively large mass ($m_{\rm{h}}\approx
200$ MeV$/c^2$ in FKP and $m_{\rm{h}}\approx 50$ MeV$/c^2$ in AMP) and
is unstable ($\tau_{\rm{h}}\approx 100$ ms in FKP and
$\tau_{\rm{h}}\approx 1$ ms in AMP). In both cases, the $\nu_{\rm{h}}$
is much heavier and shorter lived than the sterile neutrinos usually
considered in oscillation analyses \cite{Wu+2014}.  Notice that, once
they are produced in a PNS, a lifetime longer than $\tau_{\rm{h}} > 1$
s would imply that the sterile neutrinos escape the central regions of
the star but decay too far outside to have an impact on the dynamics
of the regions where the stalled shock wave is revived, whereas for
$\tau_{\rm{h}} < 10^{-7}\,$s, they are unable to scape and are just
reabsorbed by the core.  The latter possibility may, however, have
some impact on the core dynamics, since sterile neutrinos may add
another channel to homogenize the core entropy and, therefore, damp
convective instabilities.  The lifetimes proposed in the two models
evade cosmological bounds (the heavy neutrinos decay before primordial
nucleosynthesis) and do not significantly affect stars like the Sun
(they are too heavy to be produced there).  However, as discussed in
\cite{Fuller+_2009,Albertus+_2015}, these neutrinos may play a role in
the transfer of energy to the stalled shock front during a supernova
explosion. Using the numerical code \aenus\, \citep
{Obergaulinger__2008__PhD__RMHD,Just_et_al__2015__mnras__Anewmultidimensionalenergy-dependenttwo-momenttransportcodeforneutrino-hydrodynamics},
we will study quantitatively if this is the case and will estimate the
optimal value of the parameters in the models in order to facilitate
the SN explosion.

  The manuscript is structured as follows. In Section
\ref{sec:sn_theory}, we describe the models that we consider in this
work.  In Section \ref{sec:code}, we explain how to incorporate the
production and the transport of the heavy sterile neutrinos in our 1D
core-collapse supernova (CCSN) simulations. In Section
\ref{sec:results}, we present our results, studying their dependence
on the parameters of the models, and we conclude in
Sec.~\ref{sec:summary}. In the Appendix, we provide tables detailing
the production rate of sterile neutrinos as a function of temperature,
electron chemical potential and sterile mass for the AMP model as well
as discuss approximations that we made when calculating opacities of
sterile neutrinos.

\section{Heavy sterile neutrino models}
\label{sec:sn_theory}

Consider an SU(2)$_L$-singlet Dirac neutrino $\nu_{\rm{h}}$ much heavier than 
the active ones. We may denote by $N$ and
$N^c$ the two-component spinors that define $\nu_{\rm{h}}$:
\begin{equation}
\nu_{\rm h}=
\left( {\begin{array}{c}
   N \\
   \bar N^c \\
        \end{array} } \right) .
    \end{equation}
    After the breaking of the electroweak symmetry, this sterile neutrino may mix
    with an active one, $\nu$, that may correspond to a single flavor or to a combination 
    of flavors. The result is a mass eigenstate, $\nu'_{\rm{h}}=(N'\; \bar N^c)$ with 
    \begin{equation}
    N'=\cos \theta \;N + \sin \theta \;\nu\,,
    \end{equation}
    that inherits the gauge couplings of $\nu$ but suppressed by a factor of $\sin\theta$.  
    Integrating out the $W$ and $Z$ bosons, we obtain the dimension-6 operators 
\begin{eqnarray}
    -{\cal L}_{\rm eff} \supset &\,&
    {G_{\rm {F} } \sin \theta\over \sqrt{2}} \left[
      \bar f \gamma_\mu ( C_V - C_A \gamma_5) f \; 
      {\bar \nu}_{\rm{h}} \gamma^\mu ( 1 - \gamma_5 ) \nu \right. \nonumber \\
    &\,&+\left. \bar f' \gamma_\mu ( 1 - \gamma_5) f \; 
      \bar \nu_{\rm{h}} \gamma^\mu ( 1 - \gamma_5 ) \ell 
    \right]+ {\rm h.c.}\,,
 \end{eqnarray}
    where $G_{\rm F}=1.16\cdot 10^{-5}\; {\rm GeV}^{-2} (\hbar c)^3$
    is the Fermi constant
, $\gamma_\mu$ are the standard Dirac
      matrices, $\gamma_5=i\gamma_0\gamma_1\gamma_2\gamma_3$, $C_V$
      and $C_A$ are the vector and axial coupling constants, respectively,
  and
    we have dropped the prime to indicate mass eigenstates.  Furthermore, $\ell$ 
    is the charged lepton 
    belonging to the same family as
    $\nu$\maa{,} and $(f\; f')$ are standard fermions
    in the same SU(2)$_L$ doublet. These couplings imply the decays
    $\nu_{\rm{h}}\to \nu {\bar \nu} \nu$ and, if kinematically possible, 
    $\nu_{\rm{h}}\to \nu \pi^0,\, \ell^- \pi^+,\, \ell^- \ell^+\nu$, whereas the 
    dominant production channel in a PNS could be $ { \bar \nu}
    \nu\to  {\bar \nu} \nu_{\rm{h}}$ \citep{Fuller+_2009}. 
\begin{figure}[!htp]
\begin{center}
\includegraphics[angle=0,width=0.5\columnwidth]{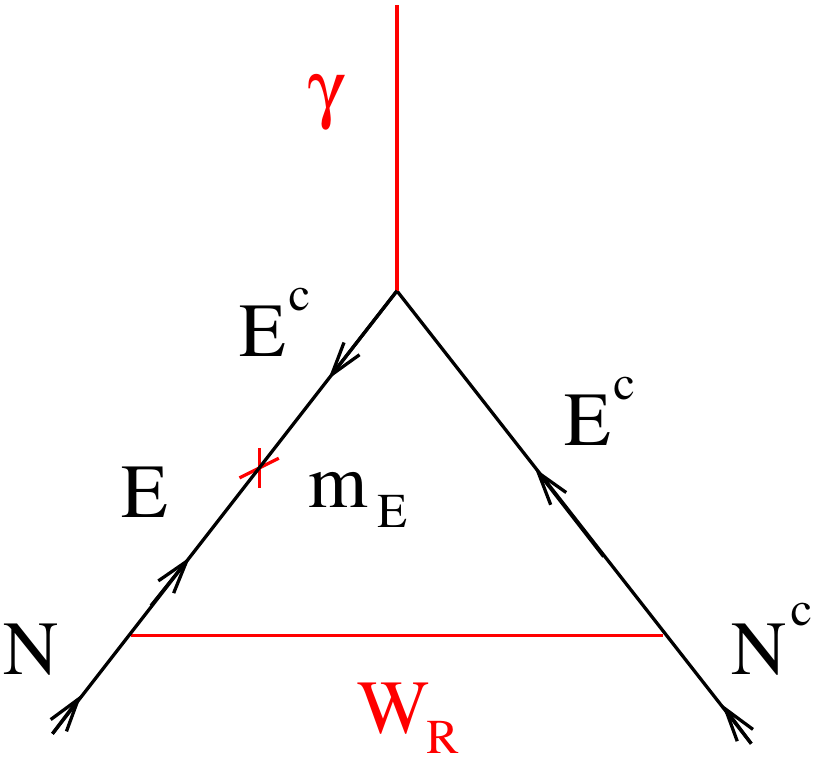} 
\end{center}
\caption{One-loop diagram generating a magnetic dipole moment $\mu_{\rm{h}}$
for $\nu_{\rm{h}}=(N\;\bar N^c)$.
Changing $N\to \tilde N$, $E\to \tilde E$ and $m_E\to m_{\tilde E E}$
we obtain an 
electromagnetic dipole transition $\mu_{\rm tr}$ between $\nu$ (the active neutrino 
mixed with $\tilde N$) and $\nu_{\rm{h}}$.
\label{fig:loop}}
\end{figure}

    The sterile neutrino $\nu_{\rm{h}}$, however, may also obtain a
    different type of couplings: dimension-5 operators generated not
    by mixing but through one-loop diagrams involving massive charged
    particles \cite{2018arXiv180303262M}. Let us be more specific.  Suppose that at the TeV
    scale, we have a left-right (L$-$R) symmetric extension of the
    SM, and that the spinors $N$ and $N^c$ come in
    SU(2)$_R$ doublets together with a charged lepton: \begin{equation} L= \left(
      {\begin{array}{c}
         N \\
         E \\
  \end{array} } \right)\,,\;\;
L^c=
\left( {\begin{array}{c}
   N^c \\
   E^c \\
        \end{array} } \right)\,.
    \end{equation}
    The breaking of the  L$-$R symmetry may then 
    result into a very massive charged lepton,
    $m_E\approx 1$ TeV$/c^2$, and a much lighter sterile neutrino, $m_{\rm{h}}=0.01$--$1\, \rm{GeV}/c^2$. In this
    case, diagrams like the one in Fig.~\ref{fig:loop} 
    will generate a magnetic dipole moment $\mu_{\rm{h}}$
    for $\nu_{\rm{h}}$ that is suppressed by only one power of the     L$-$R
    scale \cite{Bueno:2013mfa}:
    \begin{equation}
    -{\cal L}_{\rm eff}\supset
    \mu_{\rm{h}}\, \bar \nu_{\rm{h}} \sigma _{\mu \nu} \nu_{\rm{h}}\, 
\partial^\mu A^\nu \,,
\end{equation}
where $\sigma_{\mu\nu} = i [\gamma_\mu ,\gamma_\nu]/2$.
Moreover, the possible
mixing of a {\it different} sterile ${\tilde \nu}_{\rm{h}}$ (that could have a TeV$/c^2$ mass)
with the active neutrino $\nu$ 
may generate an electromagnetic dipole transition $\mu_{\rm tr}$ between 
$\nu_{\rm{h}}$ and $\nu$ of the same order even if the $\nu\, \nu_{\rm{h}}$ mixing is negligible:
\begin{equation}
-{\cal L}_{\rm eff}\supset
{1\over 2} \, \mu_{\rm tr} \,
{\overline \nu}_{\rm{h}}\, \sigma _{\mu \nu}
 \left(1-\gamma_5\right) \nu  \, \partial^\mu A^\nu + {\rm h.c.} \,.
\label{int}
\end{equation}

These couplings will introduce photon-mediated interactions of the
sterile neutrino $\nu_{\rm{h}}$ with the standard quarks and
leptons. In particular, the dominant production channel in a PNS is
expected to be $e^+ e^-\to { \bar \nu}_{\rm{h}}\nu_{\rm{h}}$, whereas
$\nu_{\rm{h}}$ will decay $\nu_{\rm{h}}\to \nu \,\gamma $
\cite{Gninenko:2010pr}.

As mentioned, for a 1--500 MeV$/c^2$  heavy neutrino, the dominant bounds 
on any model come from cosmology and from data on the 
(semi)leptonic decays of mesons (pions, kaons, heavy mesons)
and charged leptons. In all the cases
of interest its lifetime must be $\tau_{\rm{h}}<0.2$ s \cite{Dolgov:2000jw}, so that 
in the early universe sterile neutrinos decay
before primordial nucleosynthesis. 
If the lifetime is longer than $10^{-7}$ s the heavy neutrino becomes quasi-stable in 
laboratory experiments, {\it i.e.}, it tends to decay after crossing any
detector. $\nu_{\rm{h}}$ may then appear instead of the active $\nu$ in a fraction 
${\cal O}(\sin^2\theta\approx |U_{i\rm{h}}|^2)$
of meson and muon decays. Notice  that the larger the mass 
the more $\nu_{\rm{h}}$ may upset the kinematics in the process.

Masses $m_{\rm{h}}\le 30$ MeV$/c^2$ are constrained only when
$\nu_{\rm{h}}$ is mixed with the electron flavor: $\pi^+\to e^+ \nu$
puts bounds $|U_{e\rm{h}}|^2\le 10^{-6}$ at TRIUMF \cite{Britton:1992xv},
with even stronger bounds for masses up to 130 MeV$/c^2$.  At
$m_{\rm{h}} =30$--$80$ MeV$/c^2$ muon decays 
constrain  the mixing
with the $\nu_{\mu}$: $ |U_{\mu \rm{h}}|^2\le 10^{-3}$ (see discussion in
\cite{Gninenko:2010pr}).  This mixing is very constrained by recent
analyses \cite{Artamonov:2014urb,CortinaGil:2017mqf}:
$|U_{\mu \rm{h}}|^2\le 10^{-8}$ at $m_{\rm{h}}=200$--$300$ MeV$/c^2$
\cite{Artamonov:2014urb} and $|U_{\mu \rm{h}}|^2\le 2\times 10^{-7}$ at
$m_{\rm{h}}=300$--$400$ MeV$/c^2$ \cite{CortinaGil:2017mqf}.  Combined
with bounds from cosmology, these limits basically exclude the muon
possibility in the FKP  model. The bounds
on the mixing with the tau flavor, from $D_s$ meson
and $\tau$ decays, are much weaker: around $|U_{\tau \rm{h}}|^2\le 10^{-4}$
for $m_{\rm h}>160$ MeV$/c^2$ \cite{Orloff:2002de}.

As for the dimension-6 operators, for a $10^{-7}$--$0.1\,$s lifetime
the heavy neutrino becomes invisible at colliders (no bounds on
$\mu_{\rm{h}}$ and $\mu_{\rm tr}$) when the mixings vanish. In this
limit, any purely electromagnetic process giving these neutrinos will
be shadowed by an analogous $Z$-mediated process involving light
neutrinos.  In addition, the dominant decay mode
$\nu_{\rm{h}} \to \nu\gamma$ may relax the bounds on the mixings
\cite{Gninenko:2010pr} and provide an explanation for the MiniBooNE
anomaly \cite{Masip:2012ke}.

\subsection{FKP model}
The heavy neutrino proposed in \cite{Fuller+_2009} interacts with matter through $W^{\pm},Z$ boson exchange, with couplings generated through 
mixing ($\sin^2\theta < 10^{-4}$)
with the $\nu_\tau$ flavors. We set  $m_{\rm{h}} = 200\; \mev / c^2 $ as the reference value
for the mass. The main decay channel is 
$\nu_{\rm{h}} \to \nu_{\rm {\tau}} \pi^0 \to \nu_{\tau} \gamma \gamma $, 
with a lifetime 
\begin{equation}
\tau_{\rm{h}} \approx 66 \;{\rm ms}\, \left({\tento{5}{-8}\over \sin^2\theta}\right)\,
 \left( \frac{200 \; \mev }{m_{\rm{h}} c^2}  \right)^3  
 \left( \frac{0.54} {  1 -  m_{\pi}^2 / m_{\rm{h}}^2 } \right),
\label{eq:tau_fkp}
\end{equation} 
where $m_{\rm \pi} = 135\; \mev / c^2 $.  In the hot PNS
this $\nu_{\rm{h}}$ will be produced predominantly through
neutrino pair annihilation
\begin{align}  
\bar \nu_{\tau} \nu_{\tau} & \to \bar \nu_{\tau} \nu_{\rm{h}}, \nonumber \\
\bar \nu_{\mu} \nu_{\mu}  & \to \bar \nu_{\tau} \nu_{\rm{h}},  \nonumber \\
\nu_{\mu} \nu_{\tau} & \to  \nu_{\mu} \nu_{\rm{h}}, \nonumber \\
\nu_{\tau} \nu_{\tau} & \to \bar \nu_{\mu} \nu_{\rm{h}} .
\end{align} 
Other processes like pair production through nucleon-nucleon 
bremsstrahlung \citep{Bartl_et_al_2016}
will give subleading contributions due to large mass of the heavy 
neutrinos.
 A fit
of the differential luminosity in sterile neutrinos gives
\begin{equation}
  \label{eq:fkp}
  Q_{\mathrm{FKP}} \approx  \tento{3}{34}\, \frac{
    \mathrm{erg} }{ \rm{cm}^{3} \, \rm{s} }\,
  \left( \frac{ \sin^2\theta}{\tento{5}{-8} }\right)^{2}
  \left( \frac{  \kb T  }{35 \, \mathrm{MeV}}\right)^{7.2}
   e^{-\Theta_{\rm h}}\,,
\end{equation}
where $\Theta_{\rm h}\equiv \displaystyle\frac{m_{\rm{h}}c^2}{
    \kb T}$.
 We
define
\begin{equation}
\label{eq:small_q_fkp}
 q_{\mathrm{FKP}}\equiv   \left( \frac{ \sin^2\theta }{5\etoten{-8} } \right)^{2}  
\end{equation}
and rewrite Eq.~\bref{eq:fkp} as
\begin{equation}
\label{eq:fkp_with_small_q}
  Q_{\mathrm{FKP}}   \approx  \tento{3}{34}\, \frac{
    \mathrm{erg} }{ \rm{cm}^{3} \, \rm{s} }\,
  q_{\mathrm{FKP}}
  \left( \frac{  \kb T  }{35 \, \mathrm{MeV}}\right)^{7.2}
   e^{ -\maa{\Theta_{\rm h}} }\,.
\end{equation}
The parameter $q_{\mathrm{FKP}}$ can be interpreted as a production rate efficiency 
w.r.t.~the default value   $ \sin^2\theta =  5\etoten{-8}  $.

The heavy neutrinos will appear with a typical Lorentz factor of
$\gamma_{\rm{h}}\lesssim 1.5$  (see Eq.~\ref{eq:gammah})
, and their couplings to matter are so small that, once produced, they
escape the core unscattered.  Therefore, the only effect to consider
as a $\nu_{\rm{h}}$ propagates is its possible decay
$\nu_{\rm{h}} \to \nu_{\tau} \gamma \gamma$ on a timescale given by Eq.~\bref{eq:tau_fkp}.  The
initial energy carried by the heavy neutrino will be shared by the
active $\nu_{\tau}$ and the two photons (that result from the decay of
the neutral pion) which will take a fraction
\begin{equation}
  \label{eq:fkp_photons}
x_{ \gamma  \gamma }   \approx 0.5  \left( 1 + m_{\pi}^2/m_{\rm{h}}^2 \right)
\end{equation}
of energy.

We finally note that  the FKP model was proposed only in the
$m_{\rm{h}} = 145 \tto 250\; \mev / c^2 $ mass
range, and its extrapolation to higher masses is not straightforward. 
For example,
at $m_{\rm{h}} \gtrsim 300 \; \mev / c^2 $ (the heaviest neutrinos
considered in our simulations), 
 there will be new decay channels, like
\begin{equation}
  \label{eq:fkp_extra}
 \nu_{\rm{h}} \to \nu_{\tau}   \pi^+ \pi^- ,
\end{equation}
that will reduce the lifetime of the sterile neutrino. Including such 
effects is beyond
the scope of this paper, however.

\subsection{AMP model}

In the AMP model \cite{Albertus+_2015}, 
the dominant interactions of $\nu_{\rm{h}}$ with matter are 
electromagnetic. A magnetic dipole moment 
(the superindex indicates the reference value)
\begin{equation}
\label{eq:mu_h_ref} 
\mu_{\rm{ h}}^{\rm ref}= 10^{-9} \,c^{3/2}\hbar^{3/2} \, \mev^{-1}
=3.4 \cdot 10^{-9} \mu_{\rm{B}},
\end{equation}
 where $ \mu_{\rm{B}} \equiv e \hbar / (2 m_{e} c)$ is the Bohr
magneton with $e$ being the elementary charge, implies that the
dominant production channel in PNSs is
\begin{equation}
e^+ e^- \to \bar \nu_{\rm{h}} \nu_{\rm{h}}\,.
\end{equation}
The main decay mode, $\nu_{\rm{h}} \to \nu_{\mu,\tau} \gamma$, defines
a lifetime
\begin{equation}
\tau_{\rm{h}}\approx 2.6 \;{\rm ms}\, \left({\mu_{\rm tr}^{\rm ref}
\over \mu_{\rm tr}}\right)^2\, \left({50\; {\rm MeV}/c^2\over
m_h}\right)^{ 3 }
\label{eq:tau_amp}
\end{equation}
for the assumed reference value 
\begin{equation}
\label{eq:mu_tr_ref} 
\mu_{\rm tr}^{\rm ref}=3.4\cdot 10^{-11}\mu_{\rm{B}}
\end{equation}
of the dipole transition.

The coupling $\mu_{\rm tr}$ also allows active to sterile transitions
mediated by a photon and catalyzed by the presence of charged
particles in the medium: $\nu_{\mu,\tau} X \to \nu_{\rm{h}} X$ where
$X=p,e$. However, this contribution can be neglected since $\mu_{\rm
tr} < \mu_{\rm{h}}$ and in the PNS the number density of
$\nu_{\mu,\tau}$ is much smaller than that of electrons.  The
production rate of sterile neutrinos is
\begin{equation}
\label{eq:amp}
  Q_{\mathrm{AMP}} = \left( \frac{ \mu_{\rm{h}}   } { \mu_{\rm{ h}}^{\rm ref} } \right)^2  Q_{\mathrm{TAB}} ( m_{\rm{h}} , \mu_{\rm{e}}, T) ,
\end{equation}
where $ Q_{\mathrm{TAB}} ( m_{\rm{h}} , \mu_{\rm{e}}, T) $ is given by
Tab.~\ref{tab:prod_rates} (Appendix ~\ref{app:amp_prod}).  Note that the tabulated values
are more precise than the original fit deduced in Eq.~(24) of
\cite{Albertus+_2015}.
We define 
\begin{equation}
\label{eq:small_q_amp}
 q_{\mathrm{AMP}}\equiv  \left( \frac{ \mu_{\rm{h}}   } { \mu_{\rm{ h}}^{\rm ref} } \right)^2 
\end{equation}
and rewrite Eq.~\bref{eq:amp} as
\begin{equation}
\label{eq:amp_with_small_q}
  Q_{\mathrm{AMP}} =   q_{\mathrm{AMP}}  Q_{\mathrm{TAB}} ( m_{\rm{h}} , \mu_{\rm{e}}, T) .
\end{equation}
The parameter $q_{\mathrm{AMP}}$ can be interpreted as a production rate efficiency 
w.r.t.~the default value   $\mu_{\rm{h}}   = \mu_{\rm{ h}}^{\rm ref} $.

An important difference with respect to the FKP model is that 
AMP neutrinos will not leave the PNS unscattered. The reason is that their 
cross section with matter, although much smaller that the ones involving
active neutrinos, are non-negligible. In particular, the propagation of 
$\nu_{\rm{h}}$ is affected by the following three processes:
\begin{enumerate}
\item The elastic scattering  with protons. The approximate cross
  section for  this process is \cite{Albertus+_2015}
\begin{equation}
\label{eq:sigma_s_amp}
\sigma_{\rm{s}} \approx 7.5\etoten{-42}   \, \rm{cm}^{2}\, \left( \frac{ \mu_{\rm{h}}   } { \mu_{\rm{ h}}^{\rm ref}  } \right)^2\,.
\end{equation}
We can neglect the $\nu_{\rm{h}}$ scattering  off electrons as the effect is only
important for very energetic electrons, and 
due to  Fermi-blocking  such reactions will be reduced.
\item  The capture through inelastic
  collisions with charged particles:  $\nu_{\rm{h}} X\to \nu_{\mu,\tau} X$, with   $X=p,e$. 
The cross section is given by 
\begin{equation}
\label{eq:sigma_a_amp}
  \sigma_{\rm{a}}^X =   a_X \,10^{-45}   \, \rm{cm}^{2}\,
  \left( \frac{ \mu_{\text{tr}}   }{ \mu_{\text{tr}}^{\rm ref} } \right)^2 \,,
\end{equation}
with $a_{p} = 0.9$ and $a_{e} = 2.1$ for proton and electron,
respectively.
\item The decay (with the lifetime given by Eq.~\ref{eq:tau_amp}) into
  an active neutrino plus a photon,
  $\nu_{\rm{h}}\to \nu_{\mu, \tau} \gamma $ that will take a fraction
\begin{equation}
\label{eq:amp_photon}
x_{\gamma} \approx  0.5
\end{equation}  
of the energy.  %
\end{enumerate}

\section{The Code}
\label{sec:code}

\subsection{Hydrodynamics  and active neutrinos}
\label{ssec:code:MHDnu}

We added modules for evolving sterile neutrinos to a code developed
for solving the coupled system of special relativistic
(magneto-)hydrodynamics (MHD) and active-neutrino transport
\citep{Just_et_al__2015__mnras__Anewmultidimensionalenergy-dependenttwo-momenttransportcodeforneutrino-hydrodynamics}
that was used before in multidimensional supernova modelling
\citep[e.g.][]{Obergaulinger_Aloy__2017__mnras__Protomagnetarandblackholeformationinhigh-massstars,Obergaulinger_Aloy_2017JPhCS}.
The methods for solving hyperbolic equations, i.e.~high-order spatial
reconstruction, and explicit Runge-Kutta (RK) time integration, are
the basis for a very high accuracy of the solution of the MHD
equations \citep{Rembiasz_et_al_2017_num_visco}. In the simulations
presented in this paper, we use a monotonicity-preserving scheme of
the 5th order (MP5; \cite{Suresh_Huynh__1997__JCP__MP-schemes}), a
3rd-order RK time integrator, and the HLL Riemann solver
\cite{Harten_et_al__1983__SIAMReview__On_Upstream_Differencing_and_Godunov-Type_Schemes_for_Hyperbolic_Conservation_Laws}
The equations can be closed by any (tabulated) equation of state
(EOS).  Here, the EOS of
\cite{Lattimer_Swesty__1991__NuclearPhysicsA__LS-EOS} (LS-220) with an
incompressibility modulus of 220 MeV is used.  We add the fluid
self-gravity using a quasi-relativistic potential (case 'A' of
\cite{Marek_etal__2006__AA__TOV-potential}).

The active neutrinos are treated in the spectral,
i.e.~energy-dependent, hyperbolic two-moment formulation of the
transport equation, which allows for the use of the same methods as
for the MHD equations.  This scheme is based on the expansion of the
radiative intensity in its zeroth and first angular moments, i.e.~the
energy and momentum densities of the neutrinos, $E$ and $\bm{ F}$,
respectively, and closing the system of equations by a local algebraic
relation for the second moment, the radiation pressure tensor,
$P^{ij}$.  Among several possible choices for $P^{ij}$, we select the
one based on the maximum-entropy Eddington factor. Consequently, we
solve for each active neutrino species (in our case, three: $\nu_e,
\bar{\nu}_e$, and $\nu_X$ comprising all the other flavors) and for
each neutrino energy, $\varepsilon$, a system of one scalar and one
vector equation:
  \begin{align}
    \label{eq:neu-erg}
    \partial_{t} E  + \partial_{t} v_i F^i 
    +
    \nabla_i \alpha (  F^i + v^i  E )
    &
   \\    \nonumber
    -    ( \nabla_i \alpha + \dot{v}_i) \left[ \partial_{\epsilon} (\epsilon F^i) - F^i
   \right]
   & 
   \\    \nonumber
    -   \nabla_i ( \alpha v_j )
   \left[ \partial_{\epsilon} ( \epsilon P^{ij}) - P^{ij}\right]
    & =     \alpha Q_{0},
    \\
    \label{eq:neu-mom}
    \partial_{t} ( F^i + v_j P^{ij} )
     +  \nabla_j ( \alpha P^{ij} + v^j F^i) +  \dot{v}^i E & 
     \\ \nonumber
     + \alpha F^j \nabla_j v^i
    +  ( E + P^j_j) \nabla^i \alpha & 
    \\ \nonumber
    -  \partial_{\epsilon} (\epsilon P_{ij}) \dot{v}^j
    -  \alpha \partial_{\epsilon} ( \epsilon U^{ki}_j) \nabla_k v^j & 
    \\ \nonumber
    -  \partial_{\epsilon} ( \epsilon P^{ij}) \nabla_j \alpha
    & =      \alpha Q^i,
\end{align}
where $i$ and $j$ are indices which run across the three spatial
dimensions.  The equation for the moment of degree $n$ contains the
divergence of a flux involving the moment of degree $n+1$ and a term
describing the advection with the local fluid velocity, $\bm{ v}$.
This term of the equations is hyperbolic and is treated by the same
methods as the MHD equations.  Velocity and gravity are included in
the $\mathcal{O}(v/c)-\mathrm{plus}$ approximation of
\citep{Endeve_et_al__2012__ArXive-prints__ConservativeMomentEquationsforNeutrinoRadiationTransportwithLimitedRelativity}. Velocity
terms represent Doppler shifts and aberration and the effects of fluid
acceleration, while gravitational redshift and aberration are
contained in the terms involving the lapse function, $\alpha$, which
we approximate as a function of the gravitational potential as $\alpha
= \exp ( \phi/c^2)$.

  The source terms, $Q$, on the r.h.s.~of the moments equations
  describe the exchange of energy, momentum, and lepton number in
  reactions between neutrinos and matter.  Therefore, they have exact
  counterparts in the energy, momentum, and electron fraction
  equations of the MHD system.  We use a comprehensive set of
  reactions containing the absorption and emission of neutrinos by
  charged-current reactions of nucleons and nuclei and by pair
  processes (annihilation of electron-positron pairs and
  nucleon-nucleon bremsstrahlung) and scattering off nucleons, nuclei,
  and electrons/positrons (in the latter case, also accounting for
  energy transfer in non-isoenergetic scattering).  Because the
  possibly very short time scales of reactions between matter and
  neutrinos can make terms very stiff, we employ implicit time
  integrators for their solutions.

  Tests performed by \cite{Just_et_al__2015__mnras__Anewmultidimensionalenergy-dependenttwo-momenttransportcodeforneutrino-hydrodynamics} demonstrate that the code
  produces results that agree very well with the known solutions of
  simple problems and, in the case of core collapse simulations, with
  those given by state-of-the-art Boltzmann codes.

\subsection{Sterile neutrino transport}

  We adopted the same transport scheme for the sterile neutrinos.
However, lacking expressions for the dependence  of
their reactions with matter on their energy, we simplified the problem
by using the set of grey, rather than spectral, moments equations,
integrating Eqs.~(\ref{eq:neu-erg}) and (\ref{eq:neu-mom}) over
$\varepsilon$.  We note that the two-moment system is, strictly
speaking, valid only for massless particles (propagating at the speed
of light) and using it for massive sterile neutrinos is not fully
accurate and justified.  However, both kinds of simplifications should
not exceed the uncertainties related to input physics from the sterile
neutrino models.  This is why, we find using this scheme justified for
the kind of exploratory study that reported here.

  The most important ingredients that we take from the theory of
sterile neutrinos are the rates at which sterile neutrinos are
produced, their decay, and cross sections for their scattering off
matter.  In the two-moment scheme, the latter two processes contribute
to the total opacity
\begin{equation}
  \kappa = \kappa_{\mathrm{a}}  + \kappa_{\mathrm{s}}, 
\end{equation}
where ``a'' and ``s'' stand for absorption and scattering,
respectively.  The opacity has units of cm$^{-1}$, i.e., $\kappa =
\sigma n$, where $\sigma$ is a cross section and $n$ is the number
density of target particles such as nucleons, nuclei, or electrons.
In \Eqref{eq:neu-erg}, only processes that exchange energy between
neutrinos and matter appear.  The source term reads
  \begin{equation}
    \label{eq:neu-Qerg}
    Q_0 = Q_{\mathrm{p}}-c \kappa_{\mathrm{a}} E,
  \end{equation}
  where $Q_{\mathrm{P}} = Q_{\mathrm{FKP/AMP}}$ is the production term
in the FKP or AMP model given by \Eqref{eq:fkp} or
\tabref{tab:prod_rates} in Appendix \ref{app:amp_prod}, respectively.
Scattering and absorption reactions contribute to the momentum
exchange, leading to the source term
  \begin{equation}
    \label{eq:neu-Qmom}
    Q^i = -  ( \kappa_{\mathrm{a}} + \kappa_{\mathrm{s}} ) F^i.
  \end{equation}
Within this framework, we can incorporate the reaction rates of AMP
and FKP using the same numerical algorithm despite the physical
differences between both models.

In the following calculations, we  assume that $\gamma _{\rm{h}} \beta
_{\rm{h}} \approx 1$, where  $\beta _{\rm{h}} \equiv v _{\rm{h}}/c$
is the sterile neutrino velocity in terms of the speed of light. A
more detailed discussion on the validity of this approximation can be
found in Appendix \ref{app:rel}. 

In the AMP model, the absorption opacity can be estimated with the
help of \eqsref{eq:tau_amp}  and  \bref{eq:sigma_a_amp}  as
\begin{align}
\label{eq:kappa_a_amp}
  \kappa_{\rm{a}} =  & \left[ (0.9 n_{ p} + 2.1  n_{ e} ) 10^{-45} +  \tento{1.2}{-11}
    \left( \frac {m_{\rm{h}} c^2 }  { 50 \mev      } \right)^3      \right] 
  \\ \nonumber & \times \left(    \frac{\mu_{\text{tr}}    }{ \mu_{\text{tr}}^{\rm ref}  } \right)^2 \, \rm{cm}^{-1} ,
\end{align}
where $n_{\rm{e}}$ and $n_p$ are the number densities $[\rm{cm}^{-3}]$
of electrons and
protons, respectively. Furthermore,  using \Eqref{eq:sigma_s_amp}, we obtain
\begin{equation}
\kappa_{\rm{s}} =  7.5\etoten{-42}    n_{ p }\left( \frac{   \mu_{\rm{h}}    } {  \mu_{\rm{h}}^{\rm ref}   } \right)^2 \, \rm{cm}^{-1}.
\end{equation}

In  FKP model, the total  opacity is given by
\begin{equation}
  \kappa = \kappa_{\rm{a}},
\end{equation}
(i.e., there is no scattering opacity) and using
\Eqref{eq:tau_fkp}, 
we estimate
\begin{equation}
\label{eq:kappa_a_fkp}
\kappa_{\rm{a}} =  \tento{9.4}{-10} \left[ \left( \frac{ m_{\rm{h}} c^2 }{ 200 \,
      \mev } \right)^3 - 0.46  \left( \frac{ m_{\rm{h}} c^2 }{ 200\, \mev}    \right)\right]      \, \rm{cm}^{-1}  .
\end{equation}

We also assume that once sterile neutrinos decay, the energy (and
momentum) carried by created photons (given by
Eqs.~\bref{eq:amp_photon} and \bref{eq:fkp_photons} in the AMP and FKP
models, respectively) will be reabsorbed by matter and converted into
thermal energy. The energy (and momentum) carried by the active
neutrino created in the decay will be, depending on the density where
the decay occurs, carried away from the system if 
$\rho < 10^{10}\, \gccm$
 or reabsorbed by the system because of the neutrino trapping for
$\rho > 10^{12}\, \gccm$. For densities 
$10^{10}\, \gccm < \rho < 10^{12}\, \gccm $, we use a logarithmic interpolation between these
two regions. Note that this is a phenomenological prescription that we
use instead of generating another active neutrino in the code.
To test the influence of this assumption on the simulation results, we
performed two additional models A4a and A4b, which 
have all parameters like model A4, but 
threshold densities for the transition region $10^{11}\, \gccm < \rho < 10^{13}\, \gccm$,
and $10^{12}\, \gccm < \rho < 10^{14}\, \gccm$, respectively
(see Sec.~\ref{sec:results} and Tab.~\ref{tab:main_results}).

Taking into account the large uncertainties in all processes involved
and the relatively simple approach for modelling the sterile neutrinos
(non-spectral transport, no velocity terms, assumptions on the
neutrino velocity and Lorentz factor), we will consider neutrinos with
a lifetime of up to $\tau_{\mathrm{h}} = 1$\,s, {\it i.e.}, 
exceeding the bound from primordial nucleosynthesis, but still of the
same order of magnitude.

\subsection{Numerical setup}

For this study, we restrict ourselves to spherical symmetry and single
progenitor star of zero age main sequence mass $M_{\mathrm{ZAMS}} = 15
\, \msol$, namely model s15s7w2 of
\cite{Woosley_Weaver__1995__apjs__The_Evolution_and_Explosion_of_Massive_Stars.II.Explosive_Hydrodynamics_and_Nucleosynthesis}.
We deem the latter restriction justified as our goal is not arriving
at detailed predictions for specific stars.  Therefore, we selected a
standard star whose evolution is well understood as it served as a
test case in several previous studies.  The spherical symmetry
certainly limits the applicability of the simulations to real stellar
core collapse, but we accept it for such a first step towards
exploring the principle order of magnitude of the effects of sterile
neutrinos.  If indeed our study indicate interesting effects, it
should be followed up by a more thorough investigation with
multidimensional simulations.

We set up the simulations by mapping the pre-collapse model to a grid
of $608$ zones extending to $ r_{\rm max} = 10^{6}\,$km with spacing
$\Delta r = 0.0186 r + 0.2\,$km.
 This numerical resolution has been
chosen on the basis of our previous experience in the simulation of
supernova explosions with standard active neutrinos
\citep[e.g.][]{Obergaulinger_et_al__2014__mnras__Magneticfieldamplificationandmagneticallysupportedexplosionsofcollapsingnon-rotatingstellarcores,Obergaulinger_et_al__2014__ASTRONUM2013__ANewTwo-momentSchemewithAlgebraicClosureforEnergy-dependentMulti-flavorNeutrinoTransportinSupernovae,Obergaulinger_Aloy__2017__mnras__Protomagnetarandblackholeformationinhigh-massstars},
as well as a convergence study whereby one of the models incorporating
sterile neutrinos has been rerun with resolutions two and three times
larger than in our default numerical set up (see
Tab.\,\ref{tab:main_results} and the discussion in
Sec.\,\ref{sec:results}).  We evolve the core with different settings
for the sterile neutrinos through collapse up to $1\,$s after bounce.
Without sterile neutrinos, we observe the common outcome of core
collapse in spherical symmetry, viz.~the formation of a PNS and a
failure of the SN explosion as the shock wave stalls at a maximum
radius of 141 km and is never revived, eventually leading to collapse
to a black hole (BH) on much longer times scales.

\section{Results}
\label{sec:results}

\begin{figure*}
  \centering
  \includegraphics[width=0.49\linewidth]{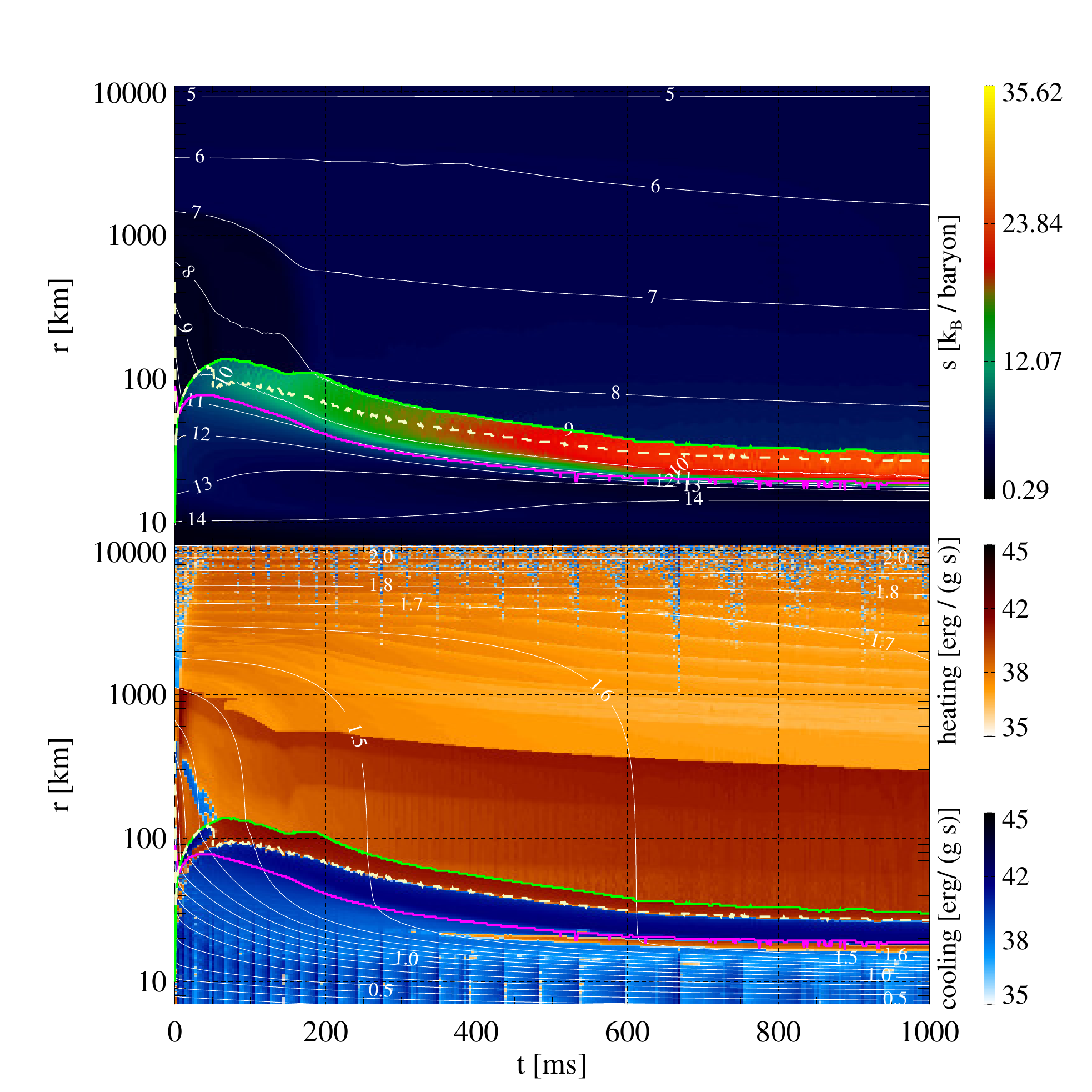} 
 \includegraphics[width=0.49\linewidth]{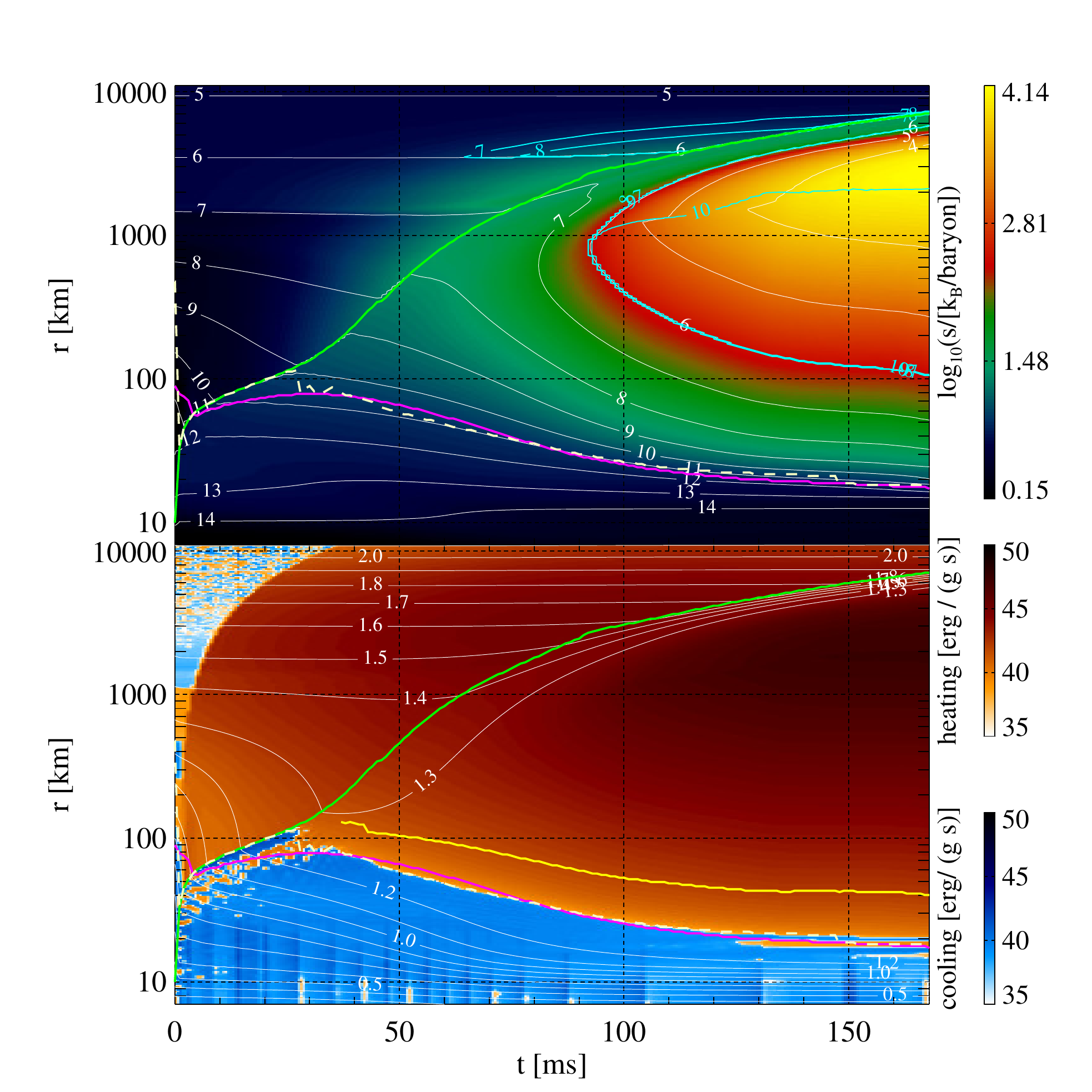} \\
 \includegraphics[width=0.49\linewidth]{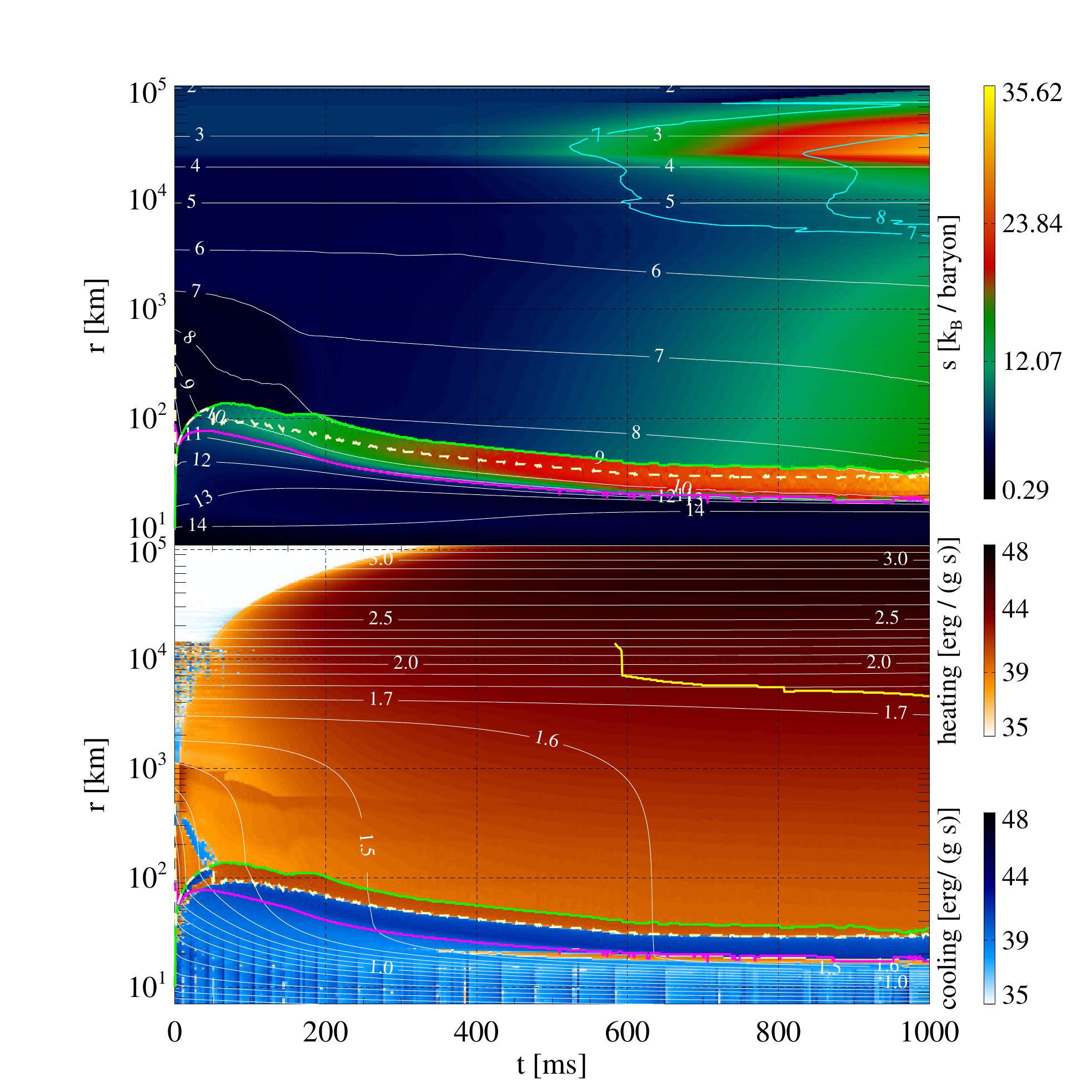}
 \includegraphics[width=0.49\linewidth]{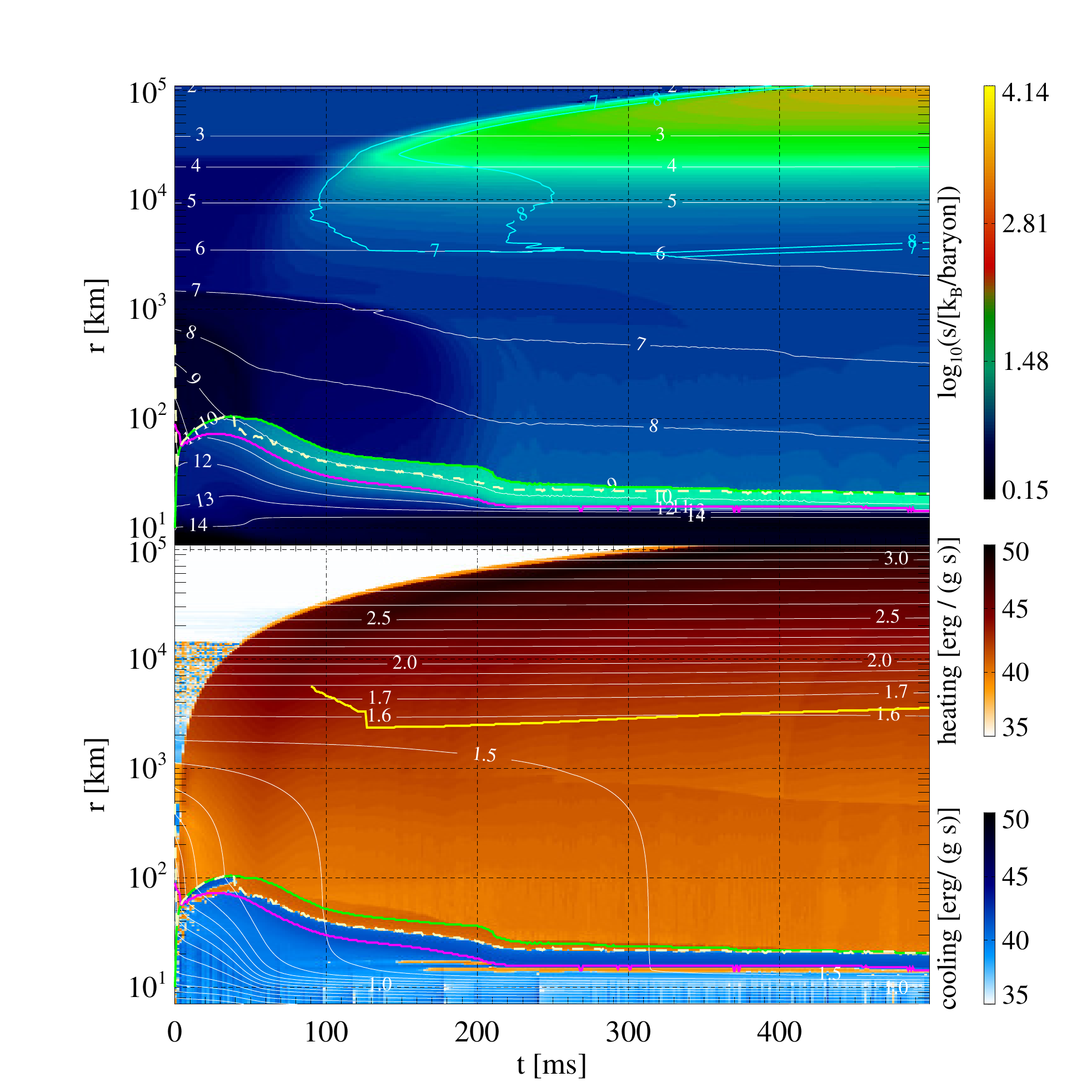}
 \caption{
   Time evolution of models R (\emph{top left}), A4 (\emph{top
     right}), F8 (\emph{bottom left}), and A8 (\emph{bottom right})
   from Tab.~\ref{tab:main_results}.  Upper subpanels: [logarithm of]
   entropy per baryon (color map), density ($\log_{10} (\rho / 1
   \gccm)$; white isocontours), positive radial velocity ($\log_{10}
   (v_r / 1\, \cm\,\sek^{-1})$; light blue
   isocontours). 
   Lower subpanels: total (i.e. active and sterile) neutrino heating
   (shades of orange) and cooling (shades of blue), enclosed mass
   (in solar masses, white isocontour), explosion radius
   (yellow line).  In both panels: shock radius (green line), neutrino
   sphere radius (proxy for the PNS radius; magenta), and gain radius
   (dashed salmon).  Note that the color scales as well as the radial
   (vertical) and temporal (horizontal) scales may vary from panel to
   panel.  }
  \label{fig:shells}
\end{figure*}

\begin{figure}
  \centering
   \includegraphics[width=0.99\columnwidth]{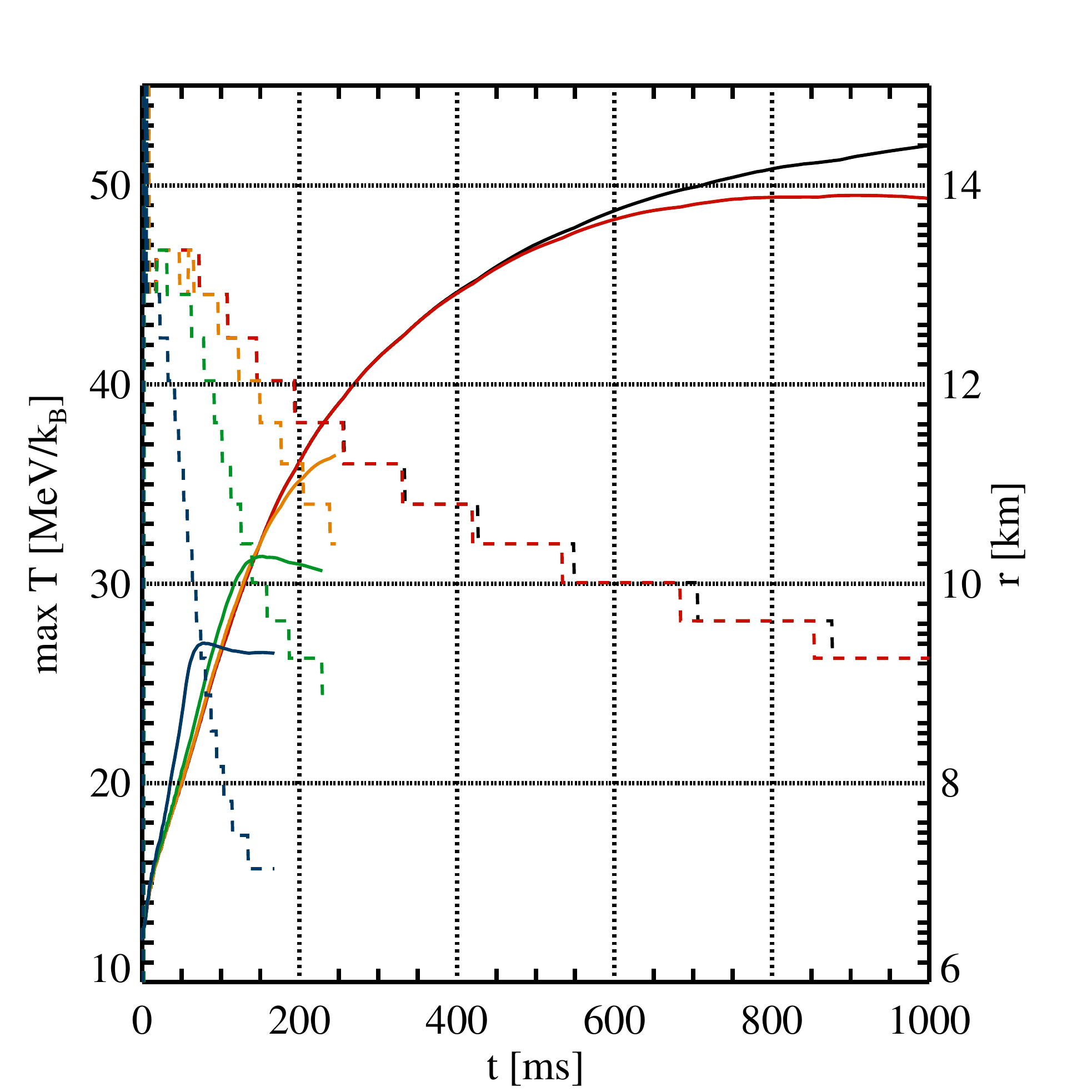}
   \caption{
     Maximum temperature (solid lines) and its location (dashed lines)
     in models: R (black), F8 (red), A23 (orange),  A16 (green), and A4 (blue).  
}
  \label{fig:max_temp}
\end{figure}

 We begin our analysis with the reference simulation (model R from
Tab.~\ref{tab:main_results}) which was run only with the three active
neutrino flavors.  For an overview of the evolution, we refer to the
upper left panel of Fig.~\ref{fig:shells} displaying entropy, neutrino
cooling and heating, and contours of the gas density as a function of
time and radius, as well as shock, gain and electron-neutrinosphere
radii.  After bounce ($t = 0\, \ms$), the shock wave of the reference
model (green solid line) stalls at a maximum radius of
$r_{\mathrm{sh}} = 141 \, \km$ at $t \approx 65 \, \ms$.  Afterwards,
neutrino heating (mapped with shades of orange in the bottom subpanel)
deposits energy in the post-shock region as well as outside the shock.
As a result, the entropy (the top subpanel) behind the shock increases
and a standard \emph{hot bubble} forms.  The active neutrino heating,
however, is insufficient to revive the shock wave.  Hence, it slowly
recedes to a radius of $r_{\mathrm{sh}} \approx 30 \, \km$ at $t =
1000 \, \ms$.  The contraction is briefly interrupted at $t \approx
160 \, \ms$ when the density of the infalling matter quickly drops
while its entropy increases as the surface of the iron core falls
through the shock.  The reduction of the ram pressure thereby produced
is, nevertheless, insufficient to spur the escape of the receding
shock, differently to what may happen in similar multidimensional
rotating and magnetized models
\cite{Obergaulinger_Aloy__2017__mnras__Protomagnetarandblackholeformationinhigh-massstars,Obergaulinger_Just_Aloy_2018}.
Taking the radius of the electron-neutrinospheres as a proxy for its
radius (magenta line), we find that the PNS contracts gradually while
it accretes matter.  We note that the evolution continues for much
longer than the $1000 \, \ms$ of post-bounce time shown here and ends
once the accretion of matter increases the PNS mass beyond the
stability limit imposed for self-gravitating objects, which for our
EOS lies above a baryonic mass of $M_{\mathrm{max}} \gtrsim2.45 \,
\msol$.  At that point, the PNS will collapse to a BH.  The maximum
temperature of the PNS (black solid line in Fig.~\ref{fig:max_temp},
which is located at a radius marked with the dashed line of the same
color) increases throughout the simulation, as more energy is provided
to the PNS through accretion of mass and contraction (which releases
gravitational energy; see the dashed lines in the figure) than
extracted by the production of active neutrinos (neutrino cooling).

\begin{table*}
  \caption[]{
Simulations performed with active neutrinos (R - \emph{reference
model}) or additionally sterile neutrinos of the AMP model (A) or the
FKP model (F).  The columns from left to right give: sterile neutrino
model, sterile neutrino mass, production efficiency w.r.t.~the default
parametres in the AMP (Eq.~\ref{eq:small_q_amp}) and the FKP
(Eq.~\ref{eq:small_q_fkp}) models, transition moment
(Eq.~\ref{eq:mu_tr_ref}), and sterile neutrino mean free path
(neglecting scattering). In the case of a successful explosion,
further columns give: explosion time, explosion radius, remnant mass
and explosion energy. These latter three quantities are determined at
the end of the simulation (given in the last column) and can change
with time.}
\footnotesize{
\begin{center}
\begin{tabular}{|c|c|c|c|c|c|c|c|c|c|c|c|c|c|c|c|c|c|c|c|c|c|c|c|c|c|c|c|c|c|c|c|c|c|c|c|c|c|c|c|c|c|c|c|c|c|c|c|}
\hline
 \#  & \pbox{0.9cm}{$m_{\rm{h}} c^2 $ \\ $[\mev]$}  & $q$ & \pbox{1.0cm}{ $\mu_{\rm{tr}} $ \\ $ [ \mu_{\rm{ tr}}^{\rm ref} ] $ }  & \pbox{0.7cm}{ $\tau_{\rm{h}} c$ \\ $[\km$] } & 
 \pbox{0.8cm}{ $t_{\rm{expl}}$ \\ $[\rm{ms}]$ }& \pbox{0.8cm}{ $r_{\rm{expl}}$ \\ $[\rm{km}]$ } & \pbox{0.7cm}{ $M_{\rm c}$ \\ $ [\Msol]$} & \pbox{0.7cm}{ $E_{\rm expl} $  \\ $[\erg$] } & \pbox{0.7cm}{ $t_{\rm end} $  \\ $[\ms$] }
\\  
  \hline  
R &   $-$  &   $-$ & $ - $  & $-$ & $-$  &   $-$  &   $-$ & $ - $    &  $1000 $
\\
\hline 
A1   &   $50$ &    $1$  &    $10$   & $7.9 $ & $-$  &   $-$  &   $-$ & $ - $    & $1000$ 
\\
A2  &   $50$ &    $1$  &    $6$ &  $22$ &  $  248 $ & $   29 $ & $ 1.40 $ & $ 2.7\etoten{51} $ & $  468 $
\\
A3  &   $50$ &    $1$  &    $3$ & $88$  & $   78 $ & $      32 $ & $ 1.22 $ & $ 1.1\etoten{52} $ & $  283 $
\\
A4  &   $50$ &    $1$  &    $1$   & $790$ & $36$ & $ 40 $ & $ 1.26 $ & $ 4.2\etoten{52} $ &  $  168 $ 
\\
A4a  &   $50$ &    $1$  &    $1$   & $790$ & $ 36 $ &  $ 40 $ & $ 1.26 $ & $ 4.3\etoten{52} $ & $  164 $ 
\\  
A4b  &   $50$ &    $1$  &    $1$   & $790$ &   $   36 $ & $      40 $ & $ 1.27 $ & $ 4.3\etoten{52} $ & $  159 $ 
\\  
A4D  &   $50$ &    $1$  &    $1$   & $790$ & $36$ &  $ 44 $ & $ 1.26 $ & $ 3.4\etoten{52} $ & $  135 $ 
\\
A4T  &   $50$ &    $1$  &    $1$   & $790$ & $37$ & $ 42 $ & $ 1.26 $ & $ 3.6\etoten{52} $ & $  136 $ 
\\
A5  &   $50$ &    $1$  &    $0.5$  &  $ 3200 $ & $38$ &   $ 46 $ & $ 1.30 $ & $ 4.6\etoten{52} $ & $  149 $ 
\\
A6  &  $50$ &    $1$  &    $0.3$ & $8800$ & $ 41 $ & $      53 $ & $ 1.33 $ & $ 3.7\etoten{52} $ & $  148 $
\\
A7   &   $50$ &    $1$  &    $ 0.1$ &  $ 7.9\etoten{4 } $  & $   65 $ & $ 1700 $ & $ 1.49 $ & $ 1.8\etoten{52} $ &   288  
\\ 
A8  &   $50$ &    $1$  &    $ 0.05$  & $3.2\etoten{5}  $ &  $  90 $  & $ 4100 $ & $ 1.65 $ & $ 1.3\etoten{52} $ & $  652 $
\\
\hline
A9  &   $50$ &    $0.3$  &    $6$ & $22$  &  $  246 $ & $      53 $ & $ 1.42 $ & $ 8.9\etoten{50} $ & $  332 $
\\
A10  &   $50$ &    $0.3$  &    $3$ & $88$  &  $   85 $ & $      29 $ & $ 1.25 $ & $ 1.3\etoten{52} $ & $  324 $
\\
A11  &   $50$ &    $0.3$  &    $1$   & $790$ & $   50 $ & $      41 $ & $ 1.29 $ & $ 4.6\etoten{52} $ & $  220 $
\\
A12  &   $50$ &    $0.3$  &    $0.3$ & $8800$ & $   59 $ & $      44 $ & $ 1.36 $ & $ 4.2\etoten{52} $ & $  208 $
\\
A13  &   $50$ &    $0.3$  &    $0.1$  & $ 7.9\etoten{4 } $ &  $   89 $ & $    1600 $ & $ 1.49 $ & $ 1.5\etoten{52} $ & $  293 $
\\
\hline
A14  &   $50$ &    $0.1$  &    $6$ & $ 22 $  &  $  253 $ & $      39 $ & $ 1.42 $ & $ 2.1\etoten{51} $ & $  439 $
\\
A15  &   $50$ &    $0.1$  &    $3$ & $88$  &   $103$  & $ 27 $ & $ 1.29 $ & $ 1.5\etoten{52} $ & $  372 $
\\
A16  &   $50$ &    $0.1$  &    $1$   & $790$ & $68$ &  $ 43 $ & $ 1.32 $ & $ 4.5\etoten{52} $ & $  229 $ 
\\
A17  &   $50$ &    $0.1$  &    $0.3$ & $8800$ & $81$ & $ 55 $ & $ 1.39 $ & $ 4.6\etoten{52} $ & $  250 $
\\
A18  &   $50$ &    $0.1$  &    $0.1$  & $ 7.9\etoten{4 } $ &  $ 117 $  &  $ 1300 $ & $ 1.49 $ & $ 1.4\etoten{52} $ & $  335 $ 
\\
\hline
A19  &   $50$ &    $3\etoten{-2}$ &    $3$ & $88$  & $  136 $ & $      26 $ & $ 1.35 $ & $ 1.5\etoten{52} $ & $  425 $
\\
A20  &   $50$ &    $3\etoten{-2}$  &    $1$   & $790$ & $   98 $ & $      37 $ & $ 1.36 $ & $ 4.8\etoten{52} $ & $  322 $
\\
A21  &   $50$ &    $3\etoten{-2}$  &    $0.3$ & $8800$ & $  112 $ & $      55 $ & $ 1.43 $ & $ 3.9\etoten{52} $ & $  292 $
\\
A22  &   $50$ &    $3\etoten{-2}$  &    $0.1$  & $ 7.9\etoten{4 } $ &  $  153 $ & $    5300 $ & $ 4.60 $ & $ 1.0\etoten{45} $ & $  152 $
\\
\hline
A23  &   $50$ &    $10^{-2 }$ &    $1$   & $790$ &  132 &  $ 41 $ & $ 1.26 $ & $ 1.9\etoten{52} $ & $  228 $
\\
A24  &   $50$ &    $10^{-2 }$ &    $0.3$   & $8800$ & 148 & $ 61 $ & $ 1.46 $ & $ 3.1\etoten{52} $ & $  336 $
\\
\hline 
  \end{tabular}
\hspace{0.5175cm} 
\begin{tabular}{|c|c|c|c|c|c|c|c|c|c|c|c|c|c|c|c|c|c|c|c|c|c|c|c|c|c|c|c|c|c|c|c|c|c|c|c|c|c|c|c|c|c|c|c|c|c|c|c|}
\hline
 \#  & \pbox{0.9cm}{$m_{\rm{h}} c^2 $ \\ $[\mev]$}  & $q$ & \pbox{1.0cm}{ $\mu_{\rm{tr}} $ \\ $ [ \mu_{\rm{ tr}}^{\rm ref} ] $ }  & \pbox{0.7cm}{ $\tau_{\rm{h}} c$ \\ $[\km$] } & 
 \pbox{0.8cm}{ $t_{\rm{expl}}$ \\ $[\rm{ms}]$ }& \pbox{0.8cm}{ $r_{\rm{expl}}$ \\ $[\rm{km}]$ } & \pbox{0.7cm}{ $M_{\rm c}$ \\ $ [\Msol]$} & \pbox{0.7cm}{ $E_{\rm expl} $  \\ $[\erg$] } & \pbox{0.7cm}{ $t_{\rm end} $  \\ $[\ms$] }
\\  
  \hline  
A25 &   $50$ &    $10^{-2 }$ &    $0.1$   & $ 7.9\etoten{4 } $ &  193 &  $ 2900 $ & $ 1.61 $ & $ 1.6\etoten{51} $ & $  265 $
\\
\hline
A26 &   $50$ &  $  3 \cdot 10^{-3 }$  &    $0.3$   &  $ 8800 $ & $  197 $ & $      26 $ & $ 1.48 $ & $ 1.4\etoten{52} $ & $  385 $ 
\\
A27 &   $50$ &   $ 3 \cdot 10^{-3 }$  &    $0.1$   &  $ 7.9\etoten{4 } $ &  $  249 $ & $    3400 $ & $ 1.65 $ & $ 4.2\etoten{51} $ & $  494 $
\\
\hline
A28 &   $50$ &    $10^{-3 }$  &    $0.1$   &  $ 7.9\etoten{4 } $ & $ 310 $ & $ 4500 $ & $ 1.73 $ & $ 4.4\etoten{51} $ & $  807 $
\\
\hline
A29  &   $50$ &    $ 0.2 $  &    $2$   & $ 200 $ &  $  67 $ & $ 34 $ & $ 1.27 $ & $ 1.9\etoten{52} $ &  $  284 $ 
\\ 
A30  &   $80$ &    $1$  &    $1$   & $190 $ &  $   73 $  &  $ 41 $ & $ 1.26 $ & $ 1.9\etoten{52} $ & $  227 $ 
\\
\hline
F1   &   $150$ &    $ 0.2$  & $-$   & $6.7\etoten{5 }$ & $ 870 $ & $ 2.6\etoten{4} $ & $ 2.55 $ & $ 6.6\etoten{49} $ & $ 1000 $
\\
F2   &   $150$ &    $1$  & $-$    & $ 1.3\etoten{5}$  & $  464 $ & $ 6300 $ & $ 1.85 $ & $ 3.0\etoten{51} $ & $ 1000 $
\\
F3   &   $150$ &    $2$  & $-$    & $6.7\etoten{4 }$ & $  382 $ & $    2000 $ & $ 1.64 $ & $ 6.6\etoten{51} $ & $  878 $
\\
F4 &   $150$ &    $6$  &    $-$   & $  2.2\etoten{4} $    & $285$  & $ 21 $ & $ 1.53 $ & $ 9.1\etoten{51} $ & $  478 $
\\
F5 &   $150$ &    $20$  &    $-$   & $ 6700  $    & $ 218 $ & $ 53 $ & $ 1.48 $ & $ 4.7\etoten{52} $ & $  397 $
\\
F6 &   $150$ &    $60$  &    $-$   & $ 2200 $    & $ 166 $ & $ 52 $ & $ 1.45 $ & $ 6.6\etoten{52} $ & $  316 $
\\
\hline 
F7   &   $200$ &    $0.2$  & $-$   & $9.8 \etoten{4}$  & $ 886 $ &  $ 2.5\etoten{4} $ & $ 2.53 $ & $ 4.0\etoten{49} $ & $  970 $
\\
F8   &   $200$ &    $1$  & $-$   & $ 2.0\etoten{4}$ & $  582 $ & $ 4600 $ & $ 1.75 $ & $ 1.3\etoten{51}$ & $ 1000 $
\\
F9   &   $200$ &    $2$  & $-$   & $9800$  & $ 493 $ & $ 210 $ & $ 1.59 $ & $ 1.9\etoten{51} $ & $  715 $ 
\\
F10   &   $200$ &    $6$  & $-$   & $3300$  & $ 369 $ & $ 65 $ & $ 1.52 $ & $ 2.1\etoten{51} $ & $  437 $ 
\\
F11   &   $200$ &    $20$  & $-$   & $980$  & $ 255 $ & $ 41 $ & $ 1.48 $ & $ 1.7\etoten{52} $ & $  441 $
\\
F12   &   $200$ &    $60$  & $-$   & $330$  & $  210 $ & $      33 $ & $ 1.46 $ & $ 3.2\etoten{52} $ & $  453 $
\\
\hline 
F13 &   $250$ &    $1$  &    $-$   & $ 7700 $  & $ - $ & $ - $ & $ - $ &   $ - $ &  $  1000$ 
\\
F14 &   $250$ &    $2$  &    $-$   & $  3800 $     &   $  761 $ &  $ 860 $ & $ 1.62 $ & $ 2.6\etoten{50} $ & $820$ 
\\
F15 &   $250$ &    $6$  &    $-$   & $  1300 $     & $  475 $ & $      22 $ & $ 1.54 $ & $ 2.7\etoten{50} $ & $  499 $
\\
F16 &   $250$ &    $20$  &    $-$   & $  390 $     &  $  349 $ & $      38 $ & $ 1.51 $ & $ 3.1\etoten{51} $ & $  452 $
\\
F17 &   $250$ &    $60$  &    $-$   & $  130 $     &  $  295 $ & $      31 $ & $ 1.49 $ & $ 9.2\etoten{51} $ & $  502 $
\\
F18 &   $250$ &    $200$  &    $-$   & $  39 $     &  $  349 $ & $      38 $ & $ 1.51 $ & $ 3.1\etoten{51} $ & $  452 $
\\
\hline 
F19 &   $300$ &    $2$  &    $-$   & $  2000 $   & $ - $ & $ - $ & $ - $ &   $ - $ &  $  957$ 
\\
F20 &   $300$ &    $6$  &    $-$   & $  660 $   & $ - $ & $ - $ & $ - $ &   $ - $ &  $  695$ 
\\
F21 &   $300$ &    $20$  &    $-$   & $  200 $    & $  527 $ & $      21 $ & $ 1.56 $ & $ 3.6\etoten{49} $ & $  535 $
\\
F22 &   $300$ &    $60$  &    $-$   & $  66 $   & $462$  & $ 31 $ & $ 1.54 $ & $ 1.2\etoten{51} $ & $  510 $
\\
F23 &   $300$ &    $200$  &    $-$   & $  20 $   & $ - $ & $ - $ & $ - $ &   $ - $ &  $  691$ 
\\
\hline 
  \end{tabular}
\label{tab:main_results}
\end{center}
}  \end{table*}

\begin{figure*}
  \centering
   \includegraphics[width=0.49\linewidth]{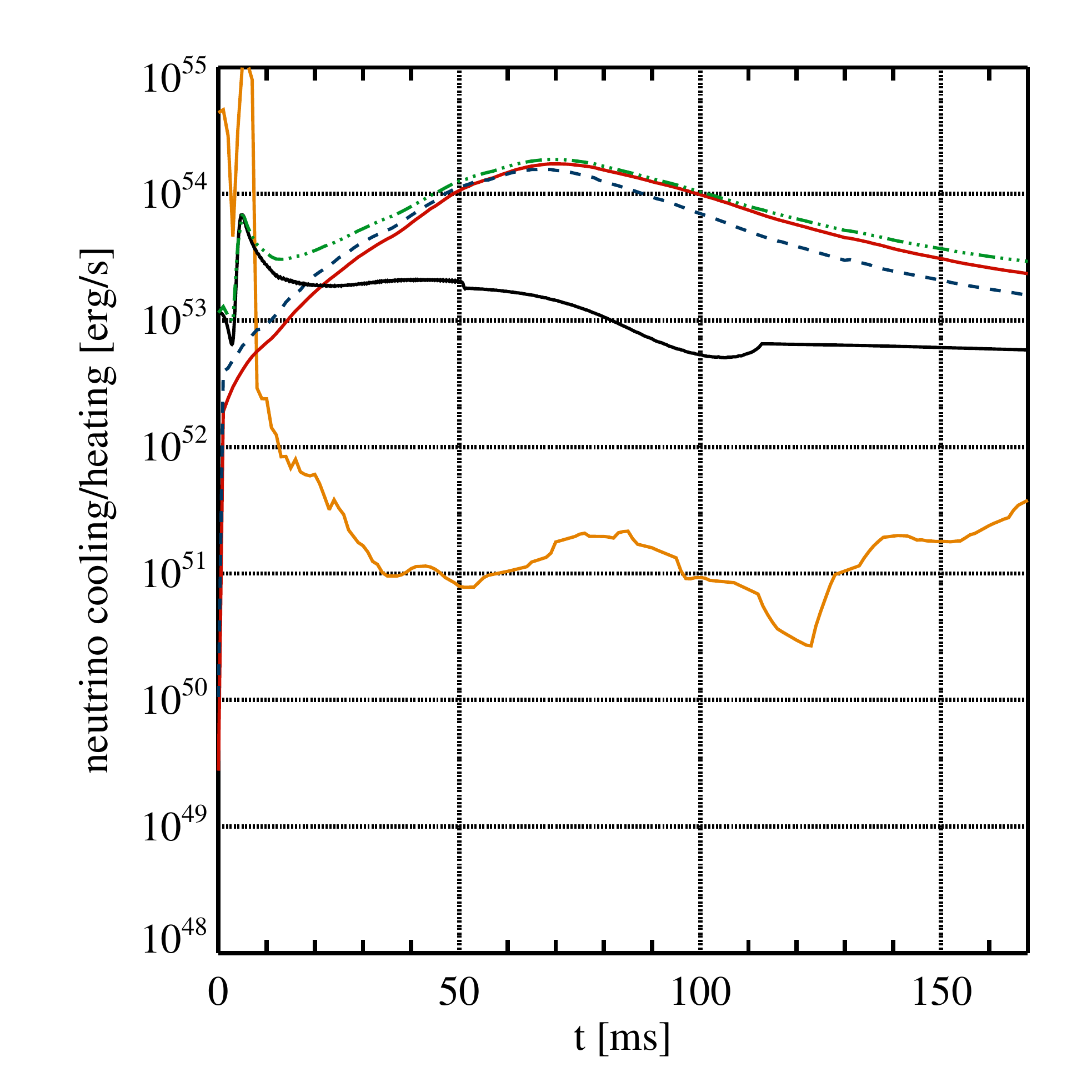}
   \includegraphics[width=0.49\linewidth]{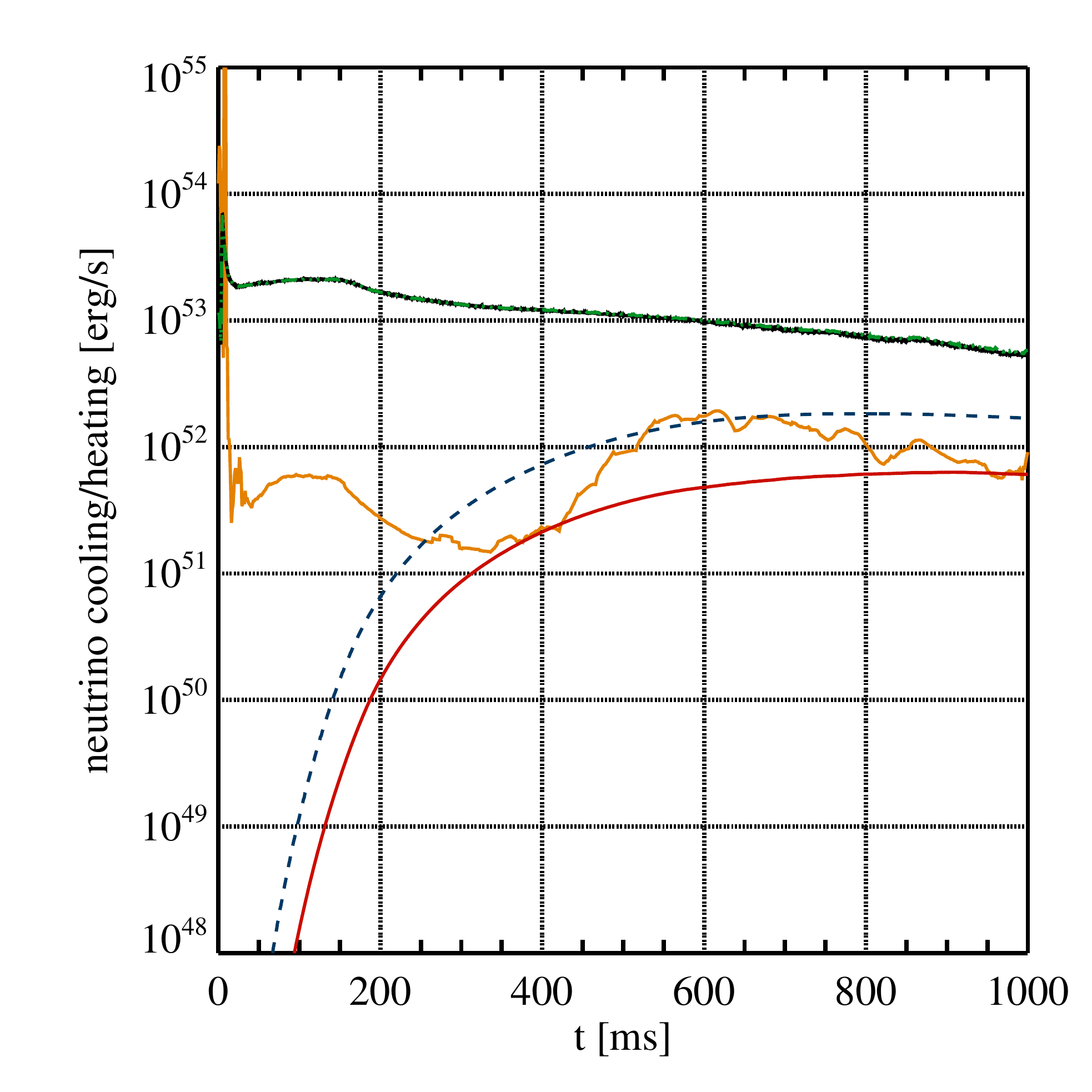}
   \caption{
     Total (volume integrated) cooling rate due to active (black) and
     sterile (blue dashed line) neutrinos, and total heating rate due to active
     (orange) and sterile (red) neutrinos in A4 (\emph{left}) and F8
     (\emph{right}) models from Tab.~\ref{tab:main_results}.
     The cooling due to active neutrinos is equivalent to the
       luminosity they would have if only SM processes were included.
       The  decays of sterile to active neutrinos add an additional
       contribution to their total luminosity, which is shown by the
       greed dash-dotted lines.}
  \label{fig:1d_hc}
\end{figure*}

\begin{figure}
  \centering
   \includegraphics[width=0.99\linewidth]{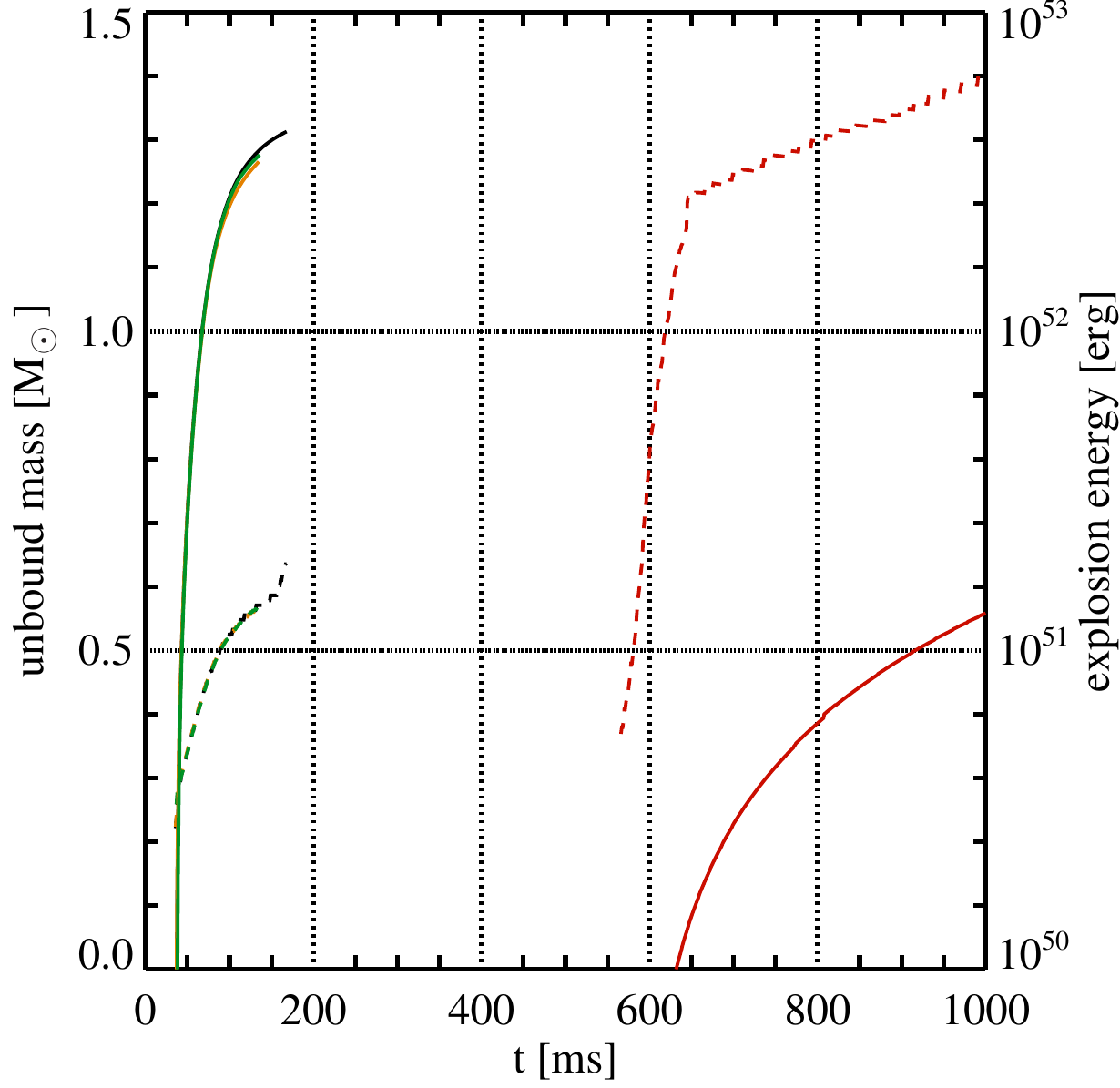} 
  \caption{
Explosion energy (Eq.~\ref{eq:e_expl};  solid lines) and unbound mass (Eq.~\ref{eq:m_expl};  dashed lines) 
in A4 (black), A4D (orange), A4T (green),  and F8 (red) models from Tab.~\ref{tab:main_results}.
Note the excellent agreement (convergence) between the models A4, A4D, and A4T  for  $ t \lesssim 130\,$ms.}
  \label{fig:1d_me}
\end{figure}

Next, we discuss the model A4 (from Tab.~\ref{tab:main_results}, see
the upper right panel of Fig.~\ref{fig:shells} as well as the left
panel of Fig.~\ref{fig:1d_hc}, and Fig.~\ref{fig:1d_me}) which was run
with (apart from the active neutrinos) sterile neutrinos of AMP with
the default parameters considered by those authors, i.e., $ m_{\rm{h}}
= 50 \, \mev/c^2$, $ \mu_{\rm h} = 10^{-9} (\hbar c)^{3/2} \,
\rm{MeV}^{-1} $ and $\mu_{\rm tr} = 10^{-11} (\hbar c)^{3/2} \,
\rm{MeV}^{-1}$.

The production of sterile neutrinos at the center of the core adds an
additional channel for cooling.  As model A4 demonstrates, the rate at
which sterile neutrinos are produced (Fig.~\ref{fig:1d_hc}; blue
dashed line), calculated as
  \begin{equation}
    \label{eq:cool_rate_tot}
\Qq^{\rm{h}}_{\mathrm{cool}}\equiv    \int Q_{\mathrm{p}} \, \der V  ,
  \end{equation}
can exceed the luminosity of active neutrinos significantly.  At
bounce, the cooling by sterile neutrinos
($\Qq^{\rm{h}}_{\mathrm{cool}} \approx 10^{50} \, \ergs$) is much
lower than that by active neutrinos ($\gtrsim 10^{53} \, \ergs$; with
black solid line, neutrino luminosity $L_{\mathrm{\nu}}$ is marked as
its proxy).  While the latter goes through the neutrino burst and then
settles to a relatively constant value $L_{\mathrm{\nu}} \sim 10^{53}
\, \ergs$, the production of sterile neutrinos is significantly
enhanced due to a rapid increase of the central temperature (from $T =
12$ to $27\, \mev /\kb$; see the solid blue line in
Fig.~\ref{fig:max_temp}).  Already at $t \approx 18\, \rm{ms}$ both
production rates become equal ($\approx 2 \cdot 10^{53} \, \ergs$),
and at $t \approx 67 \, \rm{ms}$, the production rate of sterile
neutrinos reaches its peak ($\Qq^{\rm{h}}_{\mathrm{cool}} \approx 1.6
\cdot 10^{54} \, \ergs$; $T \approx 27 \mev/\kb$) being one order of
magnitude larger than that of active neutrinos.  Afterwards, the
central temperature (and hence the production rate of sterile
neutrinos) starts to slowly drop, reaching $T \approx 26 \mev/\kb$ at
$t = 168 \, \rm{ms}$.  As a consequence, the PNS loses energy at a
higher rate and contracts faster.  This can be clearly seen from the
upper panels of Fig.~\ref{fig:shells} where the evolution of the
(electron) neutrinosphere (as a proxy for the PNS radius) is marked
with magenta for the reference (left) and A4 (right) models. At $t =
100\, \rm{ms}$, the PNS radius in the latter is $r \approx 26 \, \km$,
whereas in the former $r \approx 65 \, \km$, (the radius $r \approx 26
\, \km$ being reached only at $t \approx 400 \, \rm{ms}).$

The combination of a much more compact core together with a faster
cooling contributed by the sterile neutrinos in model A4 yields much
larger density gradients (of about one order of magnitude per grid
zone) at the surface of the PNS in this model than in the reference
one. These gradients grow with time and cause the code to fail after
about 170\,ms, because it cannot recover the thermal pressure in a
narrow region exterior to the remnant PNS. However, this is not an
insurmountable problem for the purpose of our study because the
explosion takes place much earlier than the code failure,%
\footnote{%
We note that in all models
that could not be simulated until $t = 1000\,$ms, we measure
the explosion properties (given in  Tab.~\ref{tab:main_results})
typically  $\gtrsim 10\,$ms before the code crash, so that 
they are not affected by numerical artifacts proceeding the code failure.
}
allowing us to draw the qualitative conclusion that the action of
sterile neutrinos makes viable the explosions of models which
otherwise are not exploding.  

Indeed, in order to guarantee that our results are minimally polluted
by the finite grid resolution employed, we have rerun model A4 with
double (model A4D) and triple (model A4T) number of grid zones.  We
find that the evolution of the explosion energy and of the unbound
mass converge for the first $t \approx 130\,$ms (Fig.~\ref{fig:1d_me})
with small discrepancies appearing only shortly before the code
failure.  The explosion properties ($t_{\rm expl}$, $r_{\rm expl}$ and
$E_{\rm expl}$) and the remnant mass ($M_{\rm c}$) listed in
Tab.~\ref{tab:main_results} are compatible within $\lesssim 20\% $
accuracy (the main reason of these slight deviations being different
final simulation times).

Due to their short lifetime of $2.6\,$ms and because they suffer
multiple elastic scatterings with charged matter, the sterile
neutrinos in A4 model decay within the core and deposit a large
fraction of their energy inside a radius $r =790\,$km.  Thus, the
temporal evolution of the total heating rate by sterile neutrinos
\begin{equation}
  \label{eq:heating_rate_tot}
  \Qq^{\rm{h}}_{\mathrm{heat}} \equiv  \int c \kappa_{\mathrm{a}} E  \,  \der V
\end{equation}
(solid red line in the left panel of \figref{fig:1d_hc}) is, except
for a brief time delay, the same as that of their production (dashed
blue line).  Depending sensitively on gas density and temperature, the
production term drops at the outer edge of the PNS.  Consequently, all
the gas outside the PNS is exposed to heating by the decaying sterile
neutrinos at an extraordinarily large rate exceeding a lot that due to
active neutrinos.  Thus, the shock wave does not stall its expansion
at all.

We compute the explosion energy and the ejecta mass (i.e., the unbound
mass; see Fig.~\ref{fig:1d_me}), respectively, as
\begin{align}
  \label{eq:e_expl}
 E_{\rm{expl}} & =  \int_{r \ge r_{\rm expl} }  e_{\rm tot}\,  \rho \, \der V, \\
 M_{\rm{expl}} & =  \int_{r \ge r_{\rm expl} } H( e_{\rm tot} )  \rho \, \der V,
 \label{eq:m_expl}
\end{align}
where $H$ is the Heaviside step function and
\begin{equation}
e_{\rm tot} = e_{\rm kin} + e_{\rm grav} + e_{\rm int},
\end{equation}
with $ e_{\rm kin},e_{\rm grav} ,$ and $e_{\rm int} $ being the
specific kinetic, binding (gravitational) and internal energy,
respectively.  The explosion radius, $r_{\rm expl}$, (if exists) is
defined as the innermost radius where the integral in
Eq.~\bref{eq:e_expl}, as well as both the radial velocity ($v_r(r_{\rm
expl})>0$) and total specific energy of the fluid element, $e_{\rm
tot}$, are positive.  We define the explosion time as the moment when
such an explosion radius is found.  The (time dependent) explosion
radius $r_{\rm expl} $ is marked with a yellow line in the bottom half
of the middle panel of Fig.~\ref{fig:shells}.  Note that according to
our definition, not necessarily all fluid elements located at $r
>r_{\rm expl} $, have positive $e_{\rm tot}$  (there can be
layers of the star, sufficiently far above $r_{\rm expl}$, where
$e_{\rm tot}<0$ since they have not been affected by the explosion
dynamics yet).  However, the total energy is sufficient to unbind even
the gas with a negative total energy outside $r_{\mathrm{expl}}$.
Indeed, in model A4, $M_{\rm expl} = 0.64\, \msol$ (black dashed line
in Fig.~\ref{fig:1d_me}) by the end of simulation even though the mass
contained between $r_{\rm expl} $ and $ r_{\rm max} $ is $ 3.32\,
\msol$.  For model A4, we find an explosion as early as
$t_{\mathrm{expl}} = 36\, \ms$, i.e.~we might classify the model as a
prompt explosion.

Once an explosion sets in and the gas surrounding the PNS reaches
positive (radial) velocities, the accretion ceases (see the mass shell
lines of $1.2\, \msol$ and $1.3\, \msol$ in
Fig.~\ref{fig:shells}).  Hence, the growth of the PNS mass
effectively  stops at values of $M_{\mathrm{PNS}} \simeq 1.26\,
\msol$ at $t = 168$. Thereby, this model is
unlikely to produce a BH, even though the PNS may continue to contract
as it loses internal energy by neutrino radiation.

\begin{figure}
  \centering
  \includegraphics[width=0.99\linewidth]{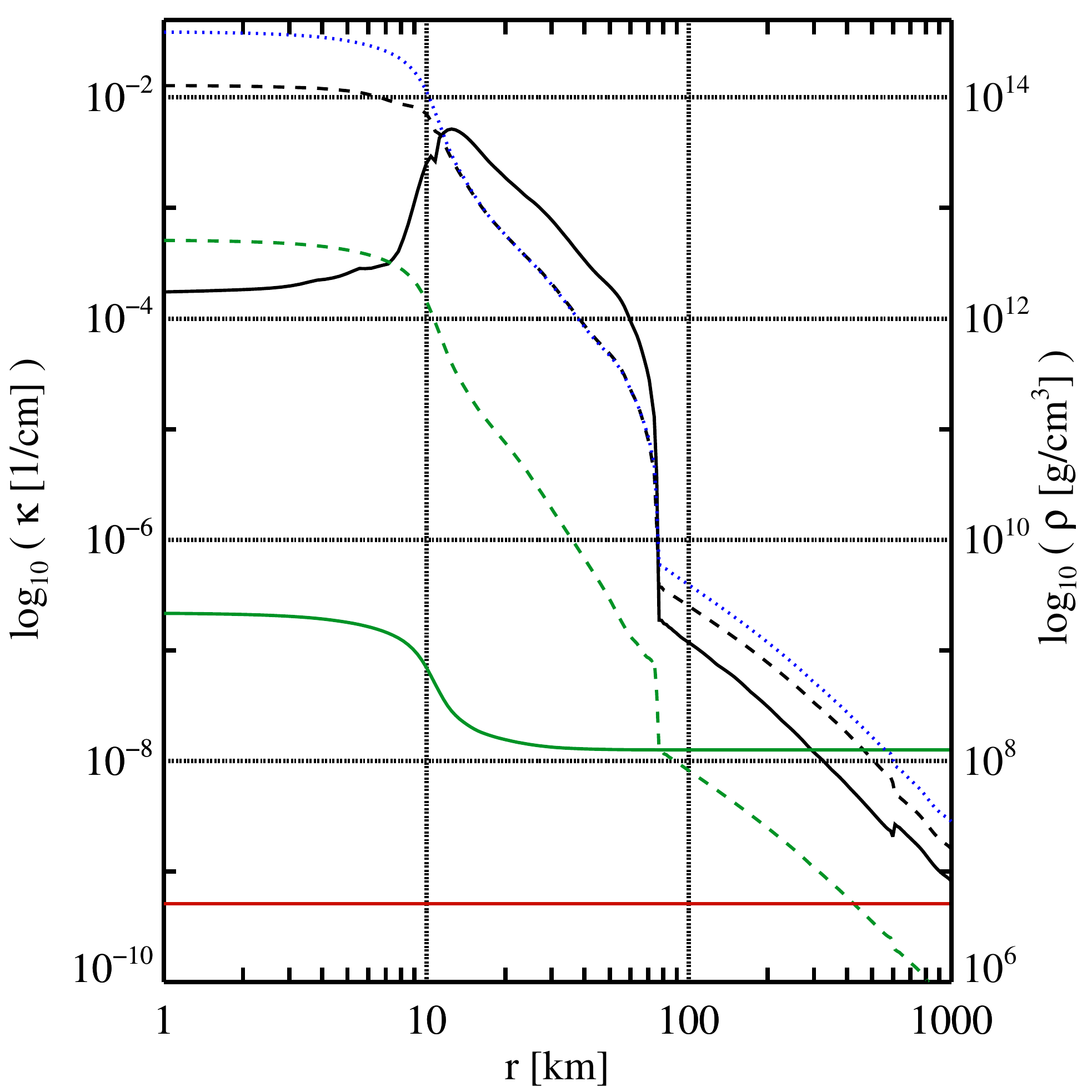}
  \caption{%
    Absorption (solid lines) and scattering (dashed lines) opacities
    of electron neutrinos with energies $75 \tto 101 \mev$ (i.e., the
    most populated energy bin; black) and of sterile neutrinos
    in simulation A4 (green) at $t = 10$\,ms post-bounce.  With red
    solid line is marked the absorption opacity of sterile neutrinos
    in model F8.  
Density profile is marked  with blue dotted line.
}
  \label{fig:kappa}
\end{figure}

The key difference between the heating by active and by sterile
neutrinos is that the latter has a contribution that does not come
from reactions between neutrinos and matter, but from the direct
decays of sterile neutrinos. Therefore, the heating rate associated to
this process does not depend on the local thermodynamics such as
density, electron fraction, and temperature, but is directly given by
the neutrino lifetime.  Hence, sterile neutrinos can efficiently heat
even cold matter at low densities where the interactions with active
neutrinos are far too infrequent for active neutrinos to be
significant.  This effect leads to the asymptotic value of the
absorption opacity for the sterile neutrinos at high radii (solid
green line in \figref{fig:kappa}), which exceeds that for active
neutrinos (solid black line) outside of a few hundred km.  It should
be furthermore noted that the strong drop of the absorption opacities
for the active neutrinos as well as the scattering opacities for both
kinds of neutrinos at the location of the shock wave ($r_{\mathrm{sh}}
\approx 70 \, \km$) is not present in the absorption opacity of
sterile neutrinos.

\begin{figure*}
  \centering
   \includegraphics[width=0.99\columnwidth]{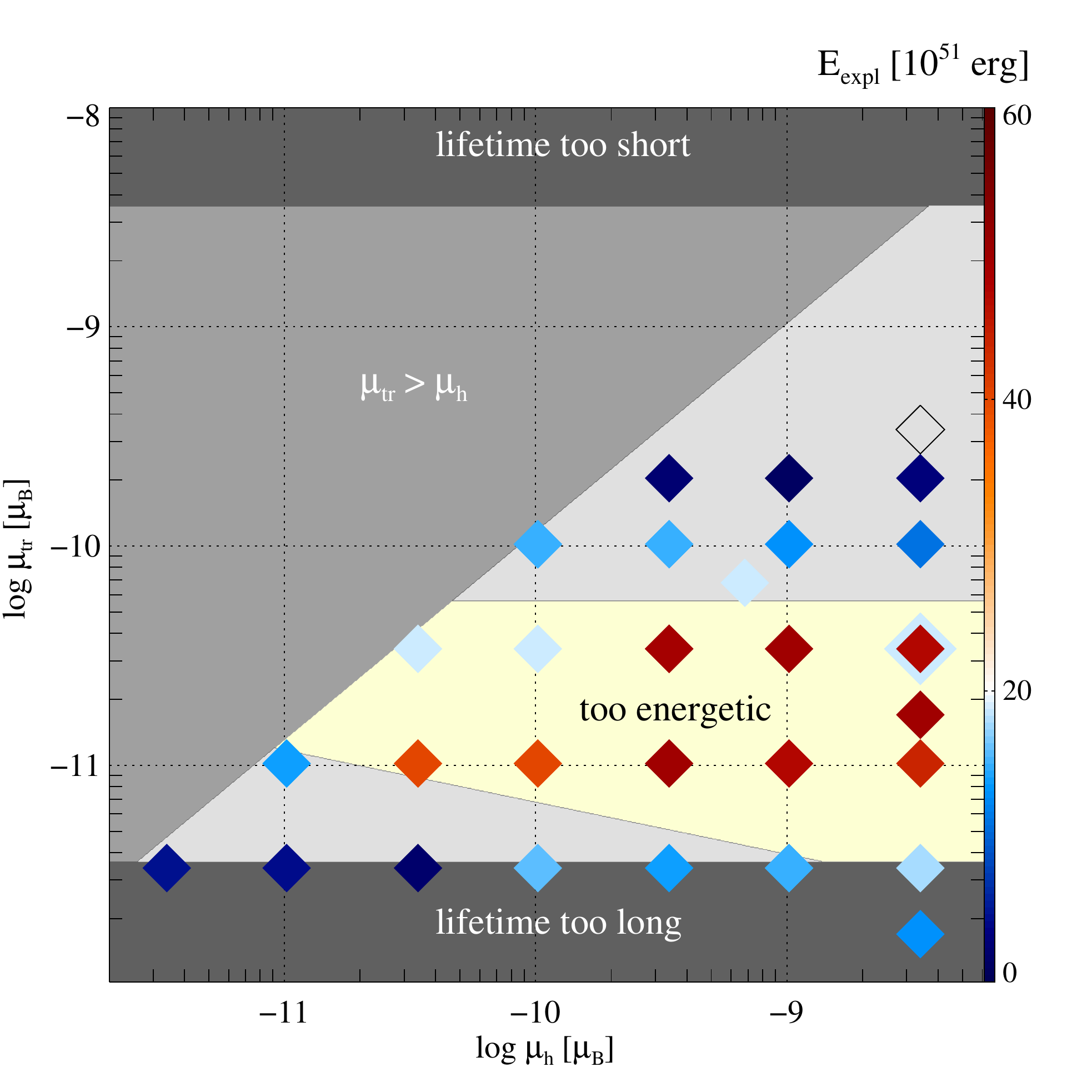}
   \includegraphics[width=0.99\columnwidth]{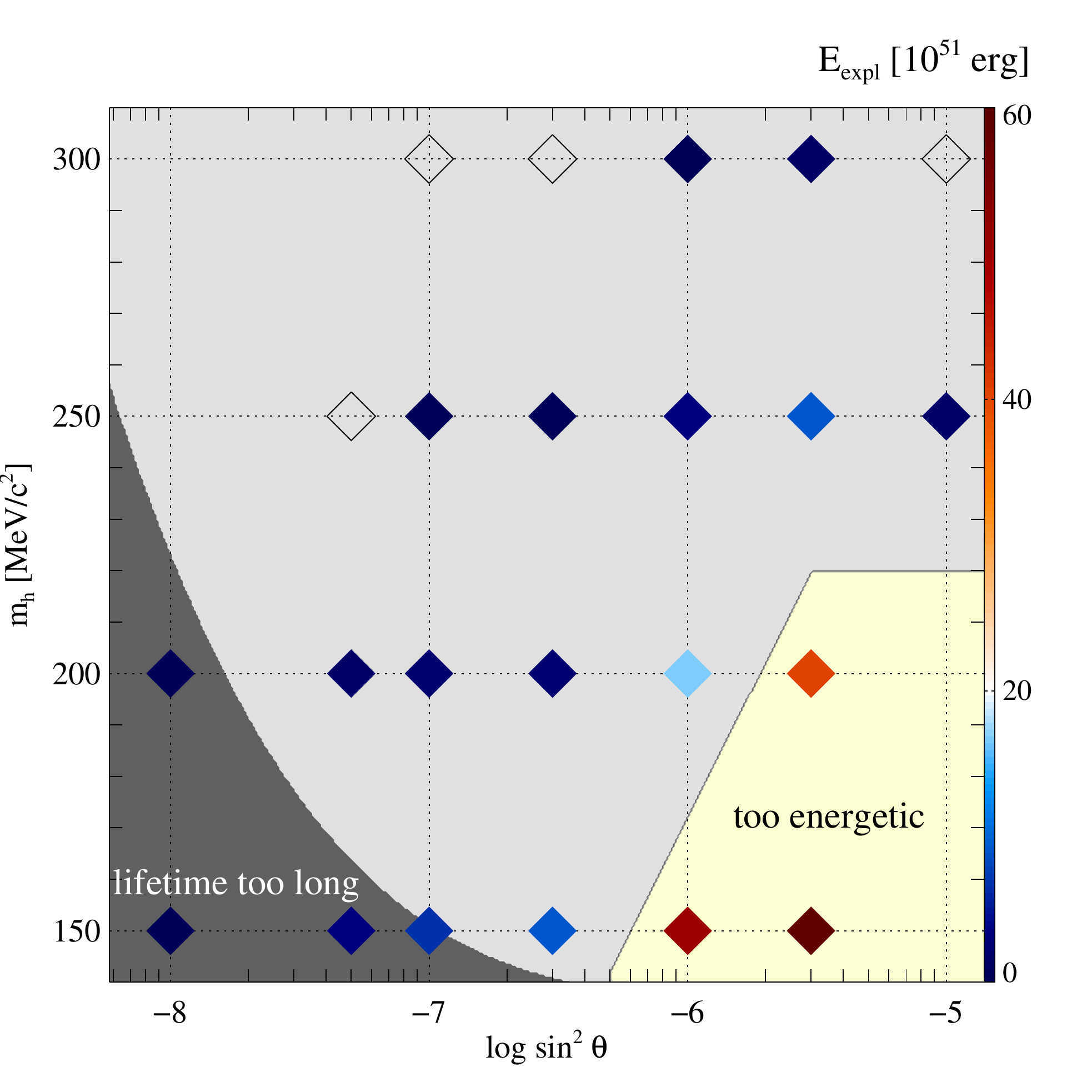}
   \caption{
Explosion energies of all AMP (left) and FKP (right)
models as a function of, respectively,  the magnetic, $\mu_{\mathrm{h}}$,
and transition moment, $\mu_{\mathrm{tr}}$ (AMP), 
and of mass, $m_{\rm{h}}$, and mixing angle $\sin^2 \theta$ (FKP)
  of the sterile neutrinos. 
In all AMP models  but A30 (with $m_{\rm{h}} =80$ MeV$/c^2$ , 
 $m_{\rm{h}} =3.4 \cdot 10^{-9} \mu_{\rm{B}},$ and  $\mu_{\mathrm{tr}} =3.4 \cdot 10^{-11} \mu_{\rm{B}}$),
sterile neutrino mass $m_{\rm{h}} =50$ MeV$/c^2$.
Non-exploding models are marked with empty diamonds.
Regions of the parameter space with an either too long (cosmological
constraint) or too short (experimental contraints)
lifetime of sterile neutrinos are marked with dark gray.
Light gray denotes region not excluded by our simulations.
Models in yellow region are too energetic ($E_{\rm expl} > 2\etoten{52} \, \rm{erg}$).
In the  gray  region of  $\mu_{\mathrm{tr}} >  \mu_{\mathrm{h}}$
(left), the production channel   $\nu X -> \nu_h X$ (not included in the
AMP model nor in our code)  is
important relative to  (or even dominant over)
$e^+ e^- -> \bar \nu_h \nu_h$\, and therefore exploring it
numerically was beyond the scope of this paper.
}
  \label{fig:kz}
\end{figure*}

  This important property of heating by sterile neutrinos causes
several peculiar differences from explosions driven by active
neutrinos.  First of all, as neutrino heating is not essentially
limited to the post-shock layer, the explosion encompasses very
quickly a very large radial range.  The shock radius (green line in
the right panel of \figref{fig:shells}) reaches a radius of
$r_{\mathrm{sh}} \approx 7040 \, \km$ at $t = 168 \, \ms$.  At this
point, $E_{\rm expl} = 4.2\etoten{52} \, \rm{erg}$ and $M_{\rm expl} =
0.64\,$\msun, and the fastest outflow velocities exceed $2\etoten{10}
\, \cms$.  All of these values are still rising by the end of the
simulation (\figref{fig:1d_me}) Remarkably, the very high explosion
energy is in the range of the most luminous hypernovae
\cite{Nomoto_et_al_2010,Mazzali+_2014,Dong+_2016}.  Secondly, as much
of the heating occurs at relatively low densities and temperatures
(note the high values of the \emph{specific} heating rate; bottom
subpanel of the upper right panel of \figref{fig:shells} at $r > 1000
\, \km$ ), the energy deposition corresponds to a very strong increase
of the gas entropy.  We find specific entropies in excess of $s >
10^{4} \, k_{\mathrm{B}} / \mathrm{baryon}$ in the ejecta and
temperatures of $ \gtrsim 0.5 \, \mev/ \kb$.  This combination of
conditions should lead to events that strongly differ from the
standard model for supernovae in terms of their observational
properties as well as their nucleosynthethic yields.  We note that a
part of the energy of the decaying sterile neutrinos should go into
active neutrinos, which will stream out from the place of their
creation almost freely.  Hence, also the light curves of active
neutrinos should be strongly increased.  We did not compute this
effect, however.

As already mentioned in Sec.~\ref{sec:code} B, 
we use a phenomenological 
prescription for the deposition of the energy and momentum of 
active neutrinos
which are created as a result of sterile neutrino 
decays instead of generating them in the code and using
the full transport equations.
We chose by default that the energy and momentum of such
neutrinos will be carried away from the system if they are created at
 $\rho < 10^{10}\, \gccm$ (because their mean free path (MFP) -
   inverse of opacity -  is
   $\gtrsim 10\,$km; see Fig.~\ref{fig:kappa})
 or reabsorbed by the system  at $\rho > 10^{12}\, \gccm$  (MFP
 $\lesssim 10\,$m).
In the transition region $10^{10}\, \gccm < \rho < 10^{12}\, \gccm$, a
logarithmic interpolation between these two scenarios is used.
To test the influence of the threshold densities,
in models A4a and A4b, we set  them to $10^{11}\, \gccm < \rho < 10^{13}\, \gccm$,
and $10^{12}\, \gccm < \rho < 10^{14}\, \gccm$, respectively.
As can be seen in  Tab.~\ref{tab:main_results}), their impact on the
explosion properties is marginal.

Next, we discuss model F8 (Tab.~\ref{tab:main_results}, see the bottom
left panel of Fig.~\ref{fig:shells} as well as the right panel of
Fig.~\ref{fig:1d_hc} and Fig.~\ref{fig:1d_me}) which was run with
sterile neutrinos of FKP with the default parameters considered by
those authors, i.e., $m_{\rm{h}} = 200 \, \mev/c^2$ and $\sin^2\theta
= \tento{5}{-8}$.  The production rate (Eq.~\ref{eq:fkp}) exhibits a
considerably different dependence on the local thermodynamic
conditions, most notably on temperature, w.r.t.~the AMP model.
Compared to model A4, it takes longer for sterile neutrinos to be
generated at significant rates (cf.~the dashed blue lines in the left
and right panels of Fig.~\ref{fig:1d_hc}).  Furthermore, their
production rate saturates at a level that is about one order of
magnitude below that of active neutrinos.  Consequently, they only
have a relatively minor influence on the evolution of the PNS whose
thermal evolution and contraction are similar to those of the
reference model.  Within the time of the simulation (i.e.,
$1000\,$ms), the radii of the two PNSs differ only slightly, as shown
by the top and bottom left panels of Fig.~\ref{fig:shells}.  The
sterile neutrinos leave an imprint in the evolution of the maximum
temperature of model F8 reaching a maximum of $T_{\mathrm{max}}
\approx 49.5 \, \mev/\kb$ at $t \approx 900 \, \ms$, while it
continuously rises to a final value of $T_{\mathrm{max}} \approx 52 \,
\mev/\kb$ at the end of the simulation of model R (see the solid red
and black lines, respectively, in Fig.~\ref{fig:max_temp}).

Another difference between A4 and F8 models is that in the latter,
sterile neutrinos have a much longer lifetime of $\tau_{\mathrm{h}} =
66\,$ms.  This leads to two effects. First, there is a noticeable
delay between the production of a sterile neutrino and its eventual
decay. Hence, its energy is temporarily unavailable to the system.
This causes a retardation of the heating by sterile neutrinos
w.r.t.~the cooling (see the right panel of Fig.~\ref{fig:1d_hc}).
Second, and more importantly, sterile neutrinos typically travel a
(much longer) distance of about $\tau_{\mathrm{h}} c \approx
\tento{2}{4} \, \mathrm{km}$ before decaying.  This is not only due to
their much lower absorption opacity (see the red line in
Fig.~\ref{fig:kappa}), but also because, unlike in the AMP model, once
produced in the center, they propagate outwards unscattered.  Hence,
heating by sterile neutrinos does not occur behind the stalled shock,
but rather outside the inner core.  Consequently, model F8 also
explodes, but does so in a very different way.  The decaying neutrinos
unbind matter at radii $r \gtrsim 4600 \, \mathrm{km}$ triggering an
explosion at time $t_{\mathrm{expl}} \approx 582 \, \mathrm{ms}$.  The
effect is most notable at several 10,000 km, where the deposition of a
large amount of energy into gas of rather low density ($\rho \lesssim
10^{4} \, \gccm$) lends itself to a strong increase of the entropy
(the bottom left panel of Fig.~\ref{fig:shells}).  After heating for
several hundreds of milliseconds, we find by the end of the simulation
an explosion energy $E_{\mathrm{expl}} \approx \tento{1.3}{51} \,
\mathrm{erg}$ carried by a mass of $M_{\mathrm{expl}} \approx 1.4 \,
\msol$ with a maximum expansion velocity of $v_{\mathrm{max}} \approx
\tento{1.9}{8} \, \cms$.  Even though these values are within the
range of standard CCSN explosions, we expect this model to produce an
electromagnetic signal significantly differing from common events
because of the relative absence of heavy elements in the ejecta. This
is because only matter from outside the iron core is unbound and the
low densities and temperatures would suppress most reactions relevant
to explosive nucleosynthesis.  We point out that the aforementioned
independence of the heating of local thermodynamic conditions is
crucial for this explosion mechanism.

So far, we have only considered sterile neutrinos of AMP and FKP with
the default parameters chosen by those authors.  However, in both
models, sterile neutrinos are characterized by, respectively, three
and two free parameters that only have rough constraints from particle
physics theory and experiments.  In AMP, these are: mass,
$m_{\rm{h}}$, magnetic dipole moment, $\mu_{\rm{ h}}$, and dipole
transition moment, $\mu_{\rm tr}$, whereas in FKP: mass, $m_{\rm{h}}$,
and mixing angle, $\sin^2\theta$.

In simulations A1--A8, we vary the transition moment of the sterile
neutrinos,$\mu_{\mathrm{tr}}$, which influences their lifetimes,
$\tau_{h} = \tento{2.6}{-5}, \dots, 1 \, \mathrm{s}$,
(Eq.~\ref{eq:tau_amp}) and the cross section on their capture through
inelastic collisions with charged particles
(Eq.~\ref{eq:sigma_a_amp}).  For the shortest lifetimes (model A1),
the sterile neutrinos travel only a few kilometeres from their
production site before decaying.  As a consequence, they do not
contribute to the energy transmission from the PNS to outer layers.
Instead, an equilibrium between matter and trapped neutrinos is
established.  The dynamics basically is the same as in the reference
model, i.e.~no explosion is launched.

Intermediate lifetimes, corresponding to typical propagation
distances of several tens to hundreds of kilometeres result in
successful shock revival. 

The most long-lived neutrinos that we investigated (models A7 and A8;
for the latter see the bottom right panel of Fig.~\ref{fig:shells})
behave similarly to model F8 discussed above.  Explosions are launched
not by reviving the stalled shock wave, but by ejecting the outer
shells of the star.  However, sterile neutrinos are produced more
efficiently than in model F8, which yields more energetic explosions.
In fact, the explosion energies in models A7 and A8 are in the range
of hypernovae.  Furthermore, the matter unbound by neutrino heating
reaches extremely high entropies in excess of $s \gtrsim 10^{4} \,
k_{\mathrm{B}} / \mathrm{baryon}$.

In simulations A4, A11, A16, A20, and A23,
 we vary the  magnetic
moment,$\mu_{\mathrm{h}}$, which influences the cross section of the
sterile neutrinos for elastic scattering with protons
(Eq.~\ref{eq:sigma_s_amp}), and, more importantly, their production
rate (Eq.~\ref{eq:amp}).  Figure \ref{fig:max_temp} displays a
positive correlation between $q_{\mathrm{AMP}}$ and the reduction of
the maximum PNS temperature reached in the course of the evolution
(w.r.t.~the reference model).  We find that the dependence of
explosion time and energy on $q_{\mathrm{AMP}}$
(Eq.~\ref{eq:small_q_amp}) is nonlinear.  
All these models (A11, A16, A20, and A23 with
$q_{\mathrm{AMP}} = 0.3,  10^{-1},  3\etoten{-2},  10^{-2},$
respectively) produce quick and strong explosions, yet in the case of model A23
 less energetic than model A4 ($q_{\mathrm{AMP}} = 1$).

We also note that the third free parameter in the AMP model, i.e.,
the sterile neutrino mass, $ m_{\rm{h}}$, does not introduce a new
kind of dynamics.  It influences both their production rate
(Tab. \ref{tab:prod_rates}) and lifetime (Eq.~\ref{eq:tau_amp}).  One
can vary either $m_{\rm{h}}$ or adequately both $\mu_{\rm{h}}$ and
$\mu_{\rm{tr}}$ achieving the same effect.  Indeed, models A29
($m_{\rm{h}} = 50 \, \mev/c^2 $, $q_{\mathrm{AMP}}= 0.2 $,
$\mu_{\rm{tr}} = 2 \mu_{\rm tr}^{\rm ref}$) and A30 ($m_{\rm{h}} = 80
\, \mev/c^2 $, $q_{\mathrm{AMP}} = 1 $, $\mu_{\rm{tr}} = \mu_{\rm
tr}^{\rm ref}$) produce essentially the same results in terms of
explosion time, radius, and energy. This demonstrates that there is a
degeneracy of the three dimensional parameter space of the AMP model
($m_{\rm{h}}$, $\mu_{\rm{h}}$, $\mu_{\rm{tr}}$).

Finally, in the left panel of Fig.\,\ref{fig:kz}, we present explosion energies of all AMP 
models A1--A30 as a function of  the magnetic, $\mu_{\mathrm{h}}$,
and transition moment , $\mu_{\mathrm{tr}}$,  of the sterile neutrinos.
Depending on theses parameters, we obtain non-exploding models (marked 
with empty diamonds), or successful explosions with energies 
which are compatible with astrophysical observations (light
gray are) or too high  ($E_{\rm expl} > 2\etoten{52} \, \rm{erg}$; yellow area).

The FKP family of models has only two free parameters, neutrino mass,
$m_{\mathrm{h}}$, and the mixing angle, $\sin^2 \theta$. Both of them
affect sterile neutrino production rate (see
Eqs.~\ref{eq:fkp}-\ref{eq:fkp_with_small_q}) and lifetime
(Eq.~\ref{eq:tau_fkp}).  Hence, this is the first difference
w.r.t.~the AMP model, where these quantities can be tweaked
independently.  Compared to the AMP model with the default parameters
(A4), the life time of the neutrinos 
in the default FKP model (F8) is long.  Thus, explosions
of the type found for model F8, i.e.~an expulsion of the outer layers
are rather common here.  Furthermore, the explosions are on average
weaker and occur later than in the AMP models.  The higher
$m_{\mathrm{h}}$, the stronger this tendency.  
Model F13 with $m_{\mathrm{h}} = 250 \, \mev / c^2$, 
and the default mixing angle (i.e., $q_{\mathrm{FKP}}=1$) fails to explode at all, and only a higher
production rate (models F14--F18) can compensate for this failure.
We note that in models F15--F18 (and others with $q_{\mathrm{FKP}} \gtrsim 6$), the explosion is triggered by the standard
shock revival mechanism.

In the right panel of Fig.~\ref{fig:kz}, where we present all FKP models,
we see that also some of them lead to too energetic explosions
 ($E_{\rm expl} > 2\etoten{52} \, \rm{erg}$). This means that also
in the FKP model, some combinations of  $m_{\mathrm{h}}$ and 
$\sin^2 \theta$ can be excluded based on our simulation results.

We finally note that the only exploding  models with 
$m_{\mathrm{h}} = 300 \, \mev / c^2$ are F21 and F22. However,
as already mentioned in Sec.~\ref{sec:sn_theory}, 
at such high masses, reactions
like in Eq.~\bref{eq:fkp_extra} should be included. This 
would  in turn lead to a further reduction of the sterile neutrino lifetimes,
and a less efficient energy transfer because of their shorter mean free path. 
Consequently, models F21 and F22 would most likely fail to yield a successful explosion.
Therefore, we may conclude that impact of the FKP neutrinos with 
$m_{\mathrm{h}} \gtrsim 300 \, \mev / c^2$ on core-collapse  would be marginal.

\section{Summary \& Conclusions}
\label{sec:summary}

We have performed 1D simulations of the core collapse of a single
progenitor star of $M_{\mathrm{ZAMS}} = 15 \, \msol$ with the two
models of \cite{Fuller+_2009} and \cite{Albertus+_2015} for sterile
neutrinos in the mass range of $50$ to  $300\,$MeV/$c^2$ that can be
produced by several different channels and decay into active neutrinos
and other particles of the SM.  In both models, the interactions of
the neutrinos depend on a few unknown free parameters (in both cases
the neutrino masses, moreover, for the FKP model the mixing angle, and
for AMP the magnetic moment as well as the transition moment), which
we varied in order to assess their influence on the dynamics of core
collapse and find potentially excluded combinations of values.  The
progenitor (model s15s7w2 of
\cite{Woosley_Weaver__1995__apjs__The_Evolution_and_Explosion_of_Massive_Stars.II.Explosive_Hydrodynamics_and_Nucleosynthesis})
is known to fail producing a successful explosion in 1D simulations
with only active neutrinos, as we verify in our reference model to
which the simulations with sterile neutrinos are compared.

  We find that the AMP model \cite{Albertus+_2015} with the default
parameters (i.e., $m_{\rm{h}} = 50\, \mev/c^2$, $\mu_{\rm{h}} =
\mu_{\rm{h}}^{\rm ref}$, and $\mu_{\rm{ tr}} = \mu_{\rm{ tr}}^{\rm
ref}$) predicts very large production rates of sterile neutrinos in
core collapse.  They are responsible for a very efficient energy
transfer (i.e., neutrino cooling and heating) from the PNS to the
post-shock matter and consequently a successful explosion. In fact,
the explosion energy, higher than $10^{52} \, \mathrm{erg}$,
corresponds to the branch of hypernovae.  However, hypernovae are only
fairly rare among stellar core collapse, and their energies clusters
at $\sim 10^{52}\,$erg with an upper limit of $\sim 2\cdot
10^{52}\,$erg\cite{Mazzali+_2014}. Taking all these facts together,
we may exclude a large part of the parameter space for the magnetic
and transition moments (see the left panel of Fig.~\ref{fig:kz}), 
including the reference values of AMP,
$\mu_{\rm{ h}}^{\rm ref} = 3.4 \cdot 10^{-9} \, \mu_{\mathrm{B}}$ and
$\mu_{\rm{ tr}}^{\rm ref} = 3.4 \cdot 10^{-11} \, \mu_{\mathrm{B}}$,
respectively.

        Lowering the magnetic moment reduces the production rate of
the sterile neutrinos and therefore their energy transfer efficiency
and leads to less energetic explosions.  Finally for $\mu_{\rm{ h} } =
10^{-2} \mu_{\rm{ h}}^{\rm ref} $ (i.e., $ q_{\mathrm{AMP}} =
10^{-4}$), no explosion occurs.

        We find that AMP sterile neutrinos with transition moments
$\mu_{\rm{ tr}} \gtrsim 5 \mu_{\rm{ tr}}^{\rm ref} $ have too short
lifetimes to change the dynamics of the core collapse in spherical
symmetry.  Due to their very short lifetime, they decay before they
leave the PNS, failing to efficiently transfer the energy necessary
for the shock revival.  We discover that sterile neutrinos with
$\mu_{\rm{ tr}} \lesssim 0.1 \mu_{\rm{ tr}}^{\rm ref} $ lead to a new
type of explosions.  Their long lifetimes allow them to travel through
the shock and only decay and release their energy in the pre-shock
matter.  Such an explosion would produce an electromagnetic signal
significantly differing from common events because of the relative
absence of heavy elements in the ejecta.

The (much better constrained) sterile neutrino mass, $m_{\rm{h}} =
50\tto 80 \, \mev/c^2$, has a secondary impact on the dynamics of the
core collapse. Moreover, we find that there is a certain degeneracy of
the three dimensional parameter space ($m_{\rm{h}}$, $\mu_{\rm{h}}$,
$\mu_{\rm{tr}}$) of the AMP model (cf.~models A29 and A30).

In the FKP model \cite{Fuller+_2009},
explosions driven by the aforementioned new mode type, that is to say,
not through shock revival, occupy a relatively large fraction of the
space of model parameters compared to the AMP model, including the
standard parameters (i.e., $m_{\rm{h}} = 200 \, \mev/c^2$ and
$\sin^2\theta = \tento{5}{-8}$).  For larger values of $\sin^2 \theta$,
however, shock revival is more likely to occur.
The maximum explosion energies can exceed 
$2 \cdot 10^{52}\,$erg, as  for the AMP models.

We conclude that sterile neutrinos of both models  can have a significant
impact on the dynamics of the core collapse.
In fact, for some paramers (allowed by the state of the art particle physics),
they can lead to too energetic explosions (see Fig.~\ref{fig:kz}).
Hence, with the help of the astrophysics data and our simulations,
it is possible to further constrain the parameter space of the hypothetical sterile neutrinos.
We plan to perform simulations with more progenitor stars  with this goal in mind.

In the spherically symmetric simulations, many
(magneto)-hydrodynamical phenomena like convection, the standing
accretion shock instability (SASI), and the magnetohydrodynamical
phenomena are suppressed.  However, they are of crucial importance for
CCSNe, as multi-dimensional models show
\cite{Janka_ARNP_2016,Janka_2017}.  They can, in particular, lead to
successful supernova explosions completely without contributions from
non-standard particle physics such as sterile neutrinos (although
still not matching all observational data).  This fact may seemingly
remove the necessity to include sterile neutrinos in supernova models.
However, we argue that, besides the possibility of constraining
neutrino models by simulations of core collapse, our models show that
sterile neutrinos can be produced at large rates even in models where
they do not cause a dramatic change in the dynamics.  In such cases,
though not the dominant component in the models, their influence might
manifest in more indirect ways.  They may affect the development of
the aforementioned instabilities by, e.g.~modifying the thermal
stratification of the core or change the cooling of the PNS over the
first few seconds after its formation.  Moreover, the entropy
stratification in the PNS may be significantly affected by the copious
numbers of sterile neutrinos effectively carrying entropy from the PNS
center to its outer layers. The most relevant consequence of such a
dynamical change could be the (partial or total) damping of the
convection close to the PNS surface. Since it is in principle possible
to observe these convective motions through the fingerprint that leave
on the gravitational wave signature (see, e.g.,
\cite{Cerda-Duran_et_al__2013__apjl__GravitationalWaveSignaturesinBlackHoleFormingCoreCollapse,Andersen_et_al_2017}),
understanding the changes induced by the action of sterile neutrinos
there is important.  Furthermore, sterile neutrinos can act as a
source of viscosity (as standard neutrinos). Thus, they may also
impact the development of the magneto-rotational instability (e.g.,
\cite{Guilet_et_al__2015__mnras__Neutrinoviscosityanddrag:impactonthemagnetorotationalinstabilityinprotoneutronstars}).
Furthermore, the production of many nuclei in supernovae is sensitive
to the detailed thermodynamic conditions of the ejecta, which might
allow to infer the existence and the properties of sterile neutrinos
from the nucleosynthetic yields of explosions.  We find it worthwhile
to investigate these effects in future
multi-dimensional simulations.

 \section*{Acknowledgements}

 MAA, MO and TR acknowledge support from the European Research Council
(grant CAMAP-259276) and from the Spanish Ministry of Economy and
Competitiveness (MINECO) and the Valencian Community grants under
grants AYA2015-66899-C2-1-P and PROMETEOII/2014-069, respectively.
The work of M.~M.~has been supported by  MINECO of Spain
(FPA2016-78220) and by Junta de Andaluc\'\i a (FQM101).  MAPG
acknowledges support from Junta de Castilla y Le\'on SA083P17 and
MINECO FIS2015-65140-P projects. MAA and MAPG thank the PHAROS COST
Action (CA16214) for partial support. M.A.~A. thanks the GWverse COST
Action (CA16104) for partial support.  We thank Meng-Ru Wu and Tobias
Fischer for valuable discussions. 
The authors thank the anonymous referee whose 
useful remarks greatly improved the quality of this manuscript.
 The computations were performed
under grants AECT-2017-2-0006, AECT-2017-3-0007 {and AECT-2018-1-0010
of the Spanish Supercomputing Network on the \textit{MareNostrum} of
the Barcelona Supercomputing Centre, and on the cluster
\textit{Lluisvives} of the Servei d'Inform\`atica of the University of
Valencia.

\appendix

\section{Production rates in AMP model}
\label{app:amp_prod}

In Table \ref{tab:prod_rates}, we provide production rates of sterile
neutrinos in the AMP model with $ \mu_{\rm{h}} = \mu_{\rm h}^{\rm
ref}=3.4 \cdot 10^{-9}\mu_{\rm{B}}$ and the lower ($m_{\rm{h}} = 50 \,
\mev/c^2 $) and upper ($m_{\rm{h}} = 80 \, \mev/c^2 $) limit for their
mass.  These tabulated values are more precise than the original fit
deduced in Eq.~(24) of \cite{Albertus+_2015}.

\begin{table*}
  \caption[]{Production rates 
$Q_{50} $  and $Q_{80} $  $[\erg \, \cm^{-3} \, \sek^{-1} ]$
of AMP sterile neutrinos with masses $ m_{\rm{h}} = 50$ and
 $ m_{\rm{h}} = 80$ MeV/$c^2$, respectively, 
as a function of temperature $ T\ [\mev/\kb]$  and chemical potential
of electrons $\mu_{ e} \ [\mev]$.}
\begin{tabular}{|c|c|c|c|c|c|c|c|c|c|c|c|c|c|c|c|c|c|c|c|}
\hline
$ T$ & $\mu_e$  & $Q_{50} $ & $ Q_{80}$ 
\\
\hline
$   5$ & $   20$ & $   1.53656\etoten{26} $ & $  5.66121\etoten{21} $
\\
$    5$ & $   40$ & $   1.36192\etoten{26} $ & $ 5.65149\etoten{21} $
\\
$    5$ & $   60$ & $   6.62017\etoten{25} $ & $ 5.38001\etoten{21} $
\\
$    5$ & $   80$ & $   1.40457\etoten{25} $ & $ 3.70775\etoten{21} $
\\
$    5$ & $   100$ & $   1.57973\etoten{24} $ & $  1.36088\etoten{21} $
\\
$    5$ & $   120$ & $   1.14701\etoten{23} $ & $  2.67659\etoten{20} $
\\
$    5$ & $   140$ & $   5.92976\etoten{21} $ & $  3.23395\etoten{19} $
\\
\hline 
$    10$ & $   20$ & $   3.35603\etoten{31} $ & $  4.6994\etoten{29} $
\\
$    10$ & $   40$ & $   2.93381\etoten{31} $ & $  4.61384\etoten{29} $
\\
$    10$ & $   60$ & $   2.0258\etoten{31} $ & $  4.23021\etoten{29} $
\\
$    10$ & $   80$ & $   1.05091\etoten{31} $ & $  3.28535\etoten{29} $
\\
$    10$ & $   100$ & $   4.24676\etoten{30} $ & $   2.01978\etoten{29} $
\\
$    10$ & $   120$ & $   1.41374\etoten{30} $ & $   9.8191\etoten{28} $
\\
$    10$ & $   140$ & $   4.06273\etoten{29} $ & $   3.90146\etoten{28} $
\\
\hline 
$    15$ & $   20$ & $   3.87518\etoten{33} $ & $  3.77445\etoten{32} $
\\
$    15$ & $   40$ & $   3.45646\etoten{33} $ & $  3.65218\etoten{32} $
\\
$    15$ & $   60$ & $   2.72116\etoten{33} $ & $  3.33863\etoten{32} $
\\
$    15$ & $   80$ & $   1.85546\etoten{33} $ & $  2.76958\etoten{32} $
\\
$    15$ & $   100$ & $   1.10604\etoten{33} $ & $   2.03238\etoten{32} $
\\
$    15$ & $   120$ & $   5.88587\etoten{32} $ & $   1.3159\etoten{32} $
\\
$    15$ & $   140$ & $   2.85587\etoten{32} $ & $   7.60523\etoten{31} $
\\
\hline
  \end{tabular}
\hspace{0.25cm}
\begin{tabular}{|c|c|c|c|c|c|c|c|c|c|c|c|c|c|c|c|c|c|c|c|}
\hline
$ T$  & $\mu_e$  & $Q_{50} $ & $ Q_{80}$ 
\\
\hline
$    20$ & $   20$ & $   5.91745\etoten{34} $ & $  1.46498\etoten{34} $
\\
$    20$ & $   40$ & $   5.40009\etoten{34} $ & $  1.40987\etoten{34} $
\\
$    20$ & $   60$ & $   4.55643\etoten{34} $ & $  1.29812\etoten{34} $
\\
$    20$ & $   80$ & $   3.52661\etoten{34} $ & $  1.12208\etoten{34} $
\\
$    20$ & $   100$ & $   2.51089\etoten{34} $ & $   8.99522\etoten{33} $
\\
$    20$ & $   120$ & $   1.65963\etoten{34} $ & $   6.67171\etoten{33} $
\\
$    20$ & $   140$ & $   1.02932\etoten{34} $ & $    4.59968\etoten{33} $
\\
\hline 
$    25$ & $   20$ & $   3.81804\etoten{35} $ & $  1.60404\etoten{35} $
\\
$    25$ & $   40$ & $   3.55223\etoten{35} $ & $  1.54426\etoten{35} $
\\
$    25$ & $   60$ & $   3.12922\etoten{35} $ & $  1.43561\etoten{35} $
\\
$    25$ & $   80$ & $   2.60148\etoten{35} $ & $  1.27651\etoten{35} $
\\
$    25$ & $   100$ & $   2.04338\etoten{35} $ & $   1.07861\etoten{35} $
\\
$    25$ & $   120$ & $   1.52328\etoten{35} $ & $   8.64511\etoten{34} $
\\
$    25$ & $   140$ & $   1.0841\etoten{35} $ & $   6.58566\etoten{34} $
\\
\hline 
$    30$ & $   20$ & $   1.56005\etoten{36} $ & $  9.09296\etoten{35} $
\\
$    30$ & $   40$ & $   1.47347\etoten{36} $ & $  8.78013\etoten{35} $
\\
$    30$ & $   60$ & $   1.33639\etoten{36} $ & $  8.24119\etoten{35} $
\\
$    30$ & $   80$ & $   1.1624\etoten{36} $ & $  7.48168\etoten{35} $
\\
$    30$ & $   100$ & $   9.70326\etoten{35} $ & $ 6.54588\etoten{35} $ 
\\
$    30$ & $   120$ & $   7.79272\etoten{35} $ & $  5.51251\etoten{35} $ 
\\
$    30$ & $   140$ & $   6.04193\etoten{35} $ & $ 4.47226\etoten{35} $
\\
\hline
  \end{tabular}
\hspace{0.25cm}
\begin{tabular}{|c|c|c|c|c|c|c|c|c|c|c|c|c|c|c|c|c|c|c|c|}
\hline
$ T$  & $\mu_e$  & $Q_{50} $ & $ Q_{80}$ 
\\
\hline
$    35$ & $   20$ & $   4.83426\etoten{36} $ & $ 3.48875\etoten{36} $
\\
$    35$ & $   40$ & $   4.61865\etoten{36} $ & $ 3.38157\etoten{36} $
\\
$    35$ & $   60$ & $   4.27645\etoten{36} $ & $  3.2018\etoten{36} $
\\
$    35$ & $   80$ & $   3.8358\etoten{36} $ & $ 2.95345\etoten{36} $
\\
$    35$ & $   100$ & $   3.33478\etoten{36} $ & $  2.64876\etoten{36} $
\\
$    35$ & $   120$ & $   2.81431\etoten{36} $ & $  2.30773\etoten{36} $
\\
$    35$ & $   140$ & $   2.31051\etoten{36} $ & $  1.95406\etoten{36} $
\\
\hline 
$    40$ & $   20$ & $   1.2473\etoten{37} $ & $ 1.03957\etoten{37} $
\\
$    40$ & $   40$ & $   1.20209\etoten{37} $ & $ 1.01149\etoten{37} $
\\
$    40$ & $   60$ & $   1.12999\etoten{37} $ & $ 9.65008\etoten{36} $
\\
$    40$ & $   80$ & $   1.03601\etoten{37} $ & $ 9.01421\etoten{36} $ 
\\
$    40$ & $   100$ & $   9.26821\etoten{36} $ & $  8.23478\etoten{36} $
\\
$    40$ & $   120$ & $   8.09845\etoten{36} $ & $  7.35349\etoten{36} $
\\
$    40$ & $   140$ & $   6.92149\etoten{36} $ & $  6.42031\etoten{36} $
\\
\hline 
$    45$ & $   20$ & $   2.82713\etoten{37} $ & $ 2.60171\etoten{37} $
\\
$    45$ & $   40$ & $   2.74285\etoten{37} $ & $ 2.54028\etoten{37} $
\\
$    45$ & $   60$ & $   2.60775\etoten{37} $ & $ 2.43927\etoten{37} $
\\
$    45$ & $   80$ & $   2.42982\etoten{37} $ & $ 2.30165\etoten{37} $
\\
$    45$ & $   100$ & $   2.21964\etoten{37} $ & $  2.13277\etoten{37} $
\\
$    45$ & $   120$ & $   1.98928\etoten{37} $ & $  1.94022\etoten{37} $
\\
$    45$ & $   140$ & $   1.7508\etoten{37} $ & $ 1.73315\etoten{37} $
\\
\hline
  \end{tabular}
\label{tab:prod_rates}
 \end{table*}

\section{Opacities} 
\label{app:rel}

Absorption opacities of the sterile neutrinos in the FKP and the AMP
models, respectively, are
 \begin{equation}
\label{eq:kappa_a_fkp_rel} \kappa_{\rm{a}} = \frac{1}{ \gamma_{\rm{h}}
\beta_{\rm{h}} \tau_{\rm{h}} c }
 \end{equation} and
 \begin{equation}
\label{eq:kappa_a_amp_rel} \kappa_{\rm{a}} = \left[ \frac{1} {
\gamma_{\rm{h}} \beta_{\rm{h}} \tau_{\rm{h}} c} + (0.9 n_{p} + 2.1
n_{e}) 10^{-45} \right] \left( \frac{ \mu_{\text{tr}} }{ \mu_{\rm
tr}^{\rm ref}} \right)^2
 \end{equation} with $\tau_{\rm{h}}$ given by Eqs.~\bref{eq:tau_amp}
and \bref{eq:tau_fkp}, respectively.  Hence, in both models, they
depend on the Lorentz factor, $\gamma_{\rm{h}}$, of the sterile
neutrino.

In the FKP model, combining Eqs.(2)-(4) from \cite{Fuller+_2009}, we
find that on average
\begin{equation} \gamma_{\rm{h}} \approx 1.3 \left( \frac{T \kb }{35
\, \mathrm{MeV}}\right)^{0.4} \left( \frac{200 \, \mathrm{MeV} }{
m_{\rm{h}} c^2 }\right).
\label{eq:gammah}
\end{equation} In the AMP model, we calculated the energy distribution
of the sterile neutrinos as a function of temperature and found that
they are produced with an average Lorentz factor
\begin{equation} \gamma_{\rm{h}} \approx 0.93 + 2.6 \frac{ T \kb
}{m_{\rm{h}} c^2 }.
\end{equation} Hence, at $T k_B < 50$ MeV, the neutrinos are in both
models mildly relativistic. Therefore, as a first approximation, in
our calculation of the opacities of sterile neutrinos, we put
$\gamma_{\rm h } \beta_{\rm h} \approx 1$ and neglected their
increased lifetimes.  In other words, we used
Eqs.~\bref{eq:kappa_a_fkp} and \bref{eq:kappa_a_amp}, instead of
Eqs.~\bref{eq:kappa_a_fkp_rel} and \bref{eq:kappa_a_amp_rel},
respectively.

\bibliographystyle{apsrev4-1} 
\bibliography{stenu}

\label{lastpage}

\end{document}